%% file: main.tex
\documentclass[sigconf]{acmart}

\renewcommand\footnotetextcopyrightpermission[1]{} 
\usepackage{xspace}
\usepackage{pifont}
\usepackage{wasysym}
\usepackage{multirow}
\usepackage{pifont}
\usepackage{subcaption}
\usepackage[normalem]{ulem}

\input{solidity-latex-highlighting.tex}

\newcommand{\ttt}[1]{\texttt{\small{#1}}}
\newcommand{\attackI}{\mbox{$\mathcal{A}_1$}\xspace}
\newcommand{\attackII}{\mbox{$\mathcal{A}_2$}\xspace}
\newcommand{\attackIII}{\mbox{$\mathcal{A}_3$}\xspace}
\newcommand{\attackIV}{\mbox{$\mathcal{A}_4$}\xspace}
\newcommand{\attackV}{\mbox{$\mathcal{A}_5$}\xspace}
\newcommand{\attackVI}{\mbox{$\mathcal{A}_6$}\xspace}

\newcommand{\msI}{\mbox{$\mathcal{M}_1$}\xspace}
\newcommand{\msII}{\mbox{$\mathcal{M}_2$}\xspace}
\newcommand{\msIII}{\mbox{$\mathcal{M}_3$}\xspace}
\newcommand{\msIV}{\mbox{$\mathcal{M}_4$}\xspace}
\newcommand{\msV}{\mbox{$\mathcal{M}_5$}\xspace}
\newcommand{\msVI}{\mbox{$\mathcal{M}_6$}\xspace}
\newcommand{\msVII}{\mbox{$\mathcal{M}_7$}\xspace}
\newcommand{\head}[1]{\textnormal{\textbf{#1}}}

\AtBeginDocument{
  \providecommand\BibTeX{{
    \normalfont B\kern-0.5em{\scshape i\kern-0.25em b}\kern-0.8em\TeX}}}

\begin{document}
\fancyhead{}

\title{Targeting the Weakest Link: Social Engineering Attacks in Ethereum Smart Contracts}

\author{Nikolay Ivanov}
\email{ivanovn1@msu.edu}
\affiliation{
  \institution{Michigan State University}
  \city{East Lansing}
  \state{MI}
  \country{USA}
  \postcode{48824}
}

\author{Jianzhi Lou}
\email{loujianz@msu.edu}
\affiliation{
  \institution{Michigan State University}
  \city{East Lansing}
  \state{MI}
  \country{USA}
  \postcode{48824}
}

\author{Ting Chen}
\email{chenting19870201@163.com}
\affiliation{
  \institution{University of Electronic Science and Technology of China}
  \city{Chengdu}
  \country{China}
}

\author{Jin Li}
\email{lijin@gzhu.edu.cn}
\affiliation{
  \institution{Guangzhou University}
  \city{Guangzhou}
  \country{China}
}

\author{Qiben Yan}
\email{qyan@msu.edu}
\affiliation{
  \institution{Michigan State University}
  \city{East Lansing}
  \state{MI}
  \country{USA}
  \postcode{48824}
}

\begin{abstract}
Ethereum holds multiple billions of U.S. dollars in the form of Ether cryptocurrency and ERC-20 tokens, with millions of deployed smart contracts algorithmically operating these funds. Unsurprisingly, the security of Ethereum smart contracts has been under rigorous scrutiny. In recent years, numerous defense tools have been developed to detect different types of smart contract code vulnerabilities. When opportunities for exploiting code vulnerabilities diminish, the attackers start resorting to social engineering attacks, which aim to influence humans --- often the weakest link in the system. The only known class of social engineering attacks in Ethereum are honeypots, which plant hidden traps for attackers attempting to exploit existing vulnerabilities, thereby targeting only a small population of potential victims.

In this work, we explore the possibility and existence of new social engineering attacks beyond smart contract honeypots. We present two novel classes of Ethereum social engineering attacks --- Address Manipulation and Homograph --- and develop six zero-day social engineering attacks. To show how the attacks can be used in popular programming patterns, we conduct a case study of five popular smart contracts with combined market capitalization exceeding \$29 billion, and integrate our attack patterns in their source codes without altering their existing functionality. Moreover, we show that these attacks remain dormant during the test phase but activate their malicious logic only at the final production deployment. We further analyze 85,656 open-source smart contracts, and discover that 1,027 of them can be used for the proposed social engineering attacks. We conduct a professional opinion survey with experts from seven smart contract auditing firms, corroborating that the exposed social engineering attacks bring a major threat to the smart contract systems.
\end{abstract}



\maketitle

\input{introduction}

\input{background}

\input{threatmodel}

\input{attacks}

\input{exploitation}

\input{evaluation}

\input{recommendations}

\input{relatedwork}

\input{conclusion}

\input{acknowledgement}

\bibliographystyle{ACM-Reference-Format}


\appendix
\input{appendix}

\end{document}

%% file: solidity-latex-highlighting.tex
\usepackage{listings, xcolor}

\definecolor{verylightgray}{rgb}{.97,.97,.97}

\lstdefinelanguage{Solidity}{
	keywords=[1]{anonymous, assembly, assert, balance, break, call, callcode, case, catch, class, constant, continue, constructor, contract, debugger, default, delegatecall, delete, do, else, emit, event, experimental, export, external, false, finally, for, function, gas, if, implements, import, in, indexed, instanceof, interface, internal, is, length, library, log0, log1, log2, log3, log4, memory, modifier, new, payable, pragma, private, protected, public, pure, push, require, return, returns, revert, selfdestruct, send, solidity, storage, struct, suicide, super, switch, then, this, throw, transfer, true, try, typeof, using, value, view, while, with, addmod, ecrecover, keccak256, mulmod, ripemd160, sha256, sha3}, 
	keywordstyle=[1]\color{blue}\bfseries,
	keywords=[2]{address, bool, byte, bytes, bytes1, bytes2, bytes3, bytes4, bytes5, bytes6, bytes7, bytes8, bytes9, bytes10, bytes11, bytes12, bytes13, bytes14, bytes15, bytes16, bytes17, bytes18, bytes19, bytes20, bytes21, bytes22, bytes23, bytes24, bytes25, bytes26, bytes27, bytes28, bytes29, bytes30, bytes31, bytes32, enum, int, int8, int16, int24, int32, int40, int48, int56, int64, int72, int80, int88, int96, int104, int112, int120, int128, int136, int144, int152, int160, int168, int176, int184, int192, int200, int208, int216, int224, int232, int240, int248, int256, mapping, string, uint, uint8, uint16, uint24, uint32, uint40, uint48, uint56, uint64, uint72, uint80, uint88, uint96, uint104, uint112, uint120, uint128, uint136, uint144, uint152, uint160, uint168, uint176, uint184, uint192, uint200, uint208, uint216, uint224, uint232, uint240, uint248, uint256, var, void, ether, finney, szabo, wei, days, hours, minutes, seconds, weeks, years},	
	keywordstyle=[2]\color{teal}\bfseries,
	keywords=[3]{block, blockhash, coinbase, difficulty, gaslimit, number, timestamp, msg, data, gas, sender, sig, value, now, tx, gasprice, origin},	
	keywordstyle=[3]\color{violet}\bfseries,
	identifierstyle=\color{black},
	sensitive=false,
	comment=[l]{//},
	morecomment=[s]{/*}{*/},
	commentstyle=\color{black}\ttfamily,
	stringstyle=\color{red}\ttfamily,
	morestring=[b]',
	morestring=[b]"
}

%% file: introduction.tex
\section{Introduction}

In one decade, the blockchain technology has emerged from a ledger of barely known cryptocurrency to an entire industry with hundreds of billions of dollars in market capitalization. A major reason of its vast expansion is the ability to support \emph{smart contracts} --- decentralized programs that can enforce execution of protocols without any third party or mutual trust. Moreover, smart contracts are used to store and transfer financial assets. For example, as of December 2020, the Tether USD smart contract had more than 2.1 million users with about \$36 billion in daily transaction volume~\cite{etherscan-tokens}.

Like any other software, smart contracts have security vulnerabilities, manifested by recent hacks with multimillion-dollar damages~\cite{mehar2019understanding,palladino2017parity}. Moreover, a recent analysis of 420 million Ethereum transactions by Zhou et al. reveals an ongoing evolution of vulnerabilities and attacks in smart contracts~\cite{zhou2020ever}. To avoid devastating consequences of smart contract hacks, a number of security auditing tools have been developed to detect smart contract vulnerabilities~\cite{tsankov2018securify,luu2016making,chen2019tokenscope,brent2018vandal}, such as reentrancy, integer overflow, etc., most of which are smart contract code vulnerabilities. However, smart contracts are designed and implemented by human developers to interact with human users, in which the human is the central component of a smart contract ecosystem. Yet, the existing smart contract security studies do not take the human factor into account. In this paper, \emph{we aim to deliver the first human-centered  study of smart contract security.}

Instead of targeting known code vulnerabilities, social engineering attacks exploit \emph{cognitive bias} of human mind. \emph{Cognitive bias} is an optimization function of the human brain that draws conclusions based on probability, expectation, previous experience, belief, or emotional response, especially when the input data is incomplete and/or decision time is limited~\cite{haselton2015evolution}. One common technique exploiting cognitive bias is \emph{visual deception}, which has been widely used in email phishing, e.g., via mimicking the appearance of a popular website~\cite{Whittaker2010LargeScaleAC} or International Domain Name (IDN) homograph attacks~\cite{holgers2006cutting}. Another aspect of cognitive bias is \emph{confirmation bias}, characterized by the rejection of evidence dissenting from the initially established belief or narrative~\cite{kappes2020confirmation}. \emph{Smart contract honeypot} is one example of confirmation bias exploitation, in which the established narrative that the smart contract is vulnerable makes even experienced hackers overlook hidden traps.

Honeypot is the only known and documented social engineering attack type in Ethereum~\cite{torres2019art}. A honeypot is a smart contract that lures a hacker into exploiting a known vulnerability, but an insidious trap in this contract turns the hacker into a victim instead. Despite being a very effective attack class, the scope of potential victims of honeypots is narrow, i.e., skillful hackers who try to steal unprotected funds.

In this work, we demonstrate that the Ethereum platform and the most popular smart contract programming language, Solidity, create a potential for evasive social engineering attacks. Social engineering attacks have been carried out across a wide spectrum of technologies, from landline phones to corporate networks. When existing software and hardware defense reduces the attack surface, the adversaries resort to exploiting human cognitive bias --- the weakest link in many security systems. To the best of our knowledge, this paper presents the first investigation of the possibility, vectors, and impact of social engineering attacks in smart contracts, as well as defense against these attacks. Specifically, we attempt to answer the following three research questions.

\noindent\textbf{RQ1: What are the Ethereum social engineering attack vectors?} We analyze the exact aspects of human cognitive bias that can be exploited to carry out social engineering attacks in smart contracts. Specifically, we discover several common misconceptions and undocumented behaviors of the Ethereum platform that create opportunities for a set of zero-day social engineering attacks.

\noindent\textbf{RQ2: Are social engineering attacks in smart contracts feasible?} Through our analysis, we identify two classes of social engineering deception --- \emph{Address Manipulation} and \emph{Homograph}. Across these two categories, we develop six social engineering attacks. By integrating the patterns of these attacks in the source codes of existing contracts with large number of users and billions of dollars in market capitalization, we further show that these attacks could potentially target a large number of victims.

\noindent\textbf{RQ3: What are the effective defenses against social engineering attacks in Ethereum?} The human is not only the main target of social engineering attacks, but also an irreplaceable element of defense against these attacks. This prompts us to develop specific security recommendations for identification and prevention of social engineering attacks by users and auditors.

In summary, we deliver the following contributions:

\begin{itemize}
    \item We identify two classes of social engineering attacks in Ethereum smart contracts, Address Manipulation and Homograph, and develop six zero-day attacks.
    
    \item We demonstrate the attacks by embedding them in source codes of five popular smart contracts with combined market capitalization of over \$29 billion, and show that the attacks have the ability to remain dormant during the testing phase and activate only after production deployment.
    
    \item We analyze 85,656 open source smart contracts and find 1,027 contracts that can be directly used for performing social engineering attacks. 
    
    \item For responsible disclosure, we contact seven smart contract security firms. The survey of experts from these firms confirms that the proposed attacks are \emph{highly likely} to be dangerous.

    \item In the spirit of open research, we make the source codes of the attack benchmark, tools, and datasets available to the public\footnote{\url{https://nick-ivanov.github.io/se-info/}}.
\end{itemize}

%% file: background.tex
\section{Background}\label{sec:background}

\noindent \textbf{Smart Contracts and EVM.}
A smart contract is a program deployed on a blockchain that provides a set of functions to be called via transactions and executed by the blockchain's virtual machine (VM). Most smart contracts are written in a high-level special-purpose programming language, such as Solidity or Vyper, and compiled into the blockchain VM bytecode. The Ethereum Virtual Machine (EVM) is the blockchain VM for executing Ethereum smart contracts.

\noindent \textbf{Externally Owned Account.}
Ethereum blockchain has two types of accounts: smart contract account and externally owned account (EOA). Both EOAs and smart contract accounts can be referenced by their 160-bit public addresses. EOAs can be used to call the functions of smart contracts via signed transactions.

\noindent \textbf{ERC-20 Tokens.}
ERC-20 is the most popular standard for implementing fungible tokens\footnote{Each fungible token has the same value and does not possess any special characteristics compared with other tokens of the same type.} in Ethereum smart contracts. Some of the most traded alternative cryptocurrencies (altcoins) are ERC-20-compatible smart contracts deployed on Ethereum Mainnet, such as ChainLink and Binance Coin. The ERC-20 standard defines an interface that a smart contract should implement in order to become an ERC-20 token to interact with ERC-20-compliant clients\footnote{https://eips.ethereum.org/EIPS/eip-20}.

\noindent \textbf{OpenZeppelin Contracts.}
\emph{OpenZeppelin Contracts} is a library of smart contracts that have been extensively tested for adherence to the best security practices. These smart contracts are considered to be the de-facto standardized implementations of popular smart contract code patterns. The OpenZeppelin project provides a rich codebase for ERC-20 token developers\footnote{https://openzeppelin.com/contracts/}. 

\noindent \textbf{EIP-55 Checksums.}
Developers of blockchain clients use checksums for validating public addresses. A checksum is a digital fingerprint of an address to ensure its validity and correctness. In Ethereum, the checksum is embedded in the address by capitalizing certain hexadecimal letters, as described in the EIP-55 standard\footnote{https://eips.ethereum.org/EIPS/eip-55}. Specifically, if the $i^{\text{th}}$ hexadecimal digit of Keccak256 hash digest of the EIP-55 address string is $\geq$ 8, the $i^{\text{th}}$ hexadecimal digit of the address is capitalized. The accuracy of EIP-55 error checking is nearly 99.986\%~\cite{antonopoulos2018mastering}.

\noindent\textbf{Smart Contract Addresses.}
A smart contract address in Ethereum is generated using the deterministic function\footnote{An implementation of this function, named \texttt{generateAddress}, can be found at https://github.com/ethereumjs/ethereumjs-util.} $\chi(A_d,\eta)$, where $A_d$ is the public address of the account deploying the contract, and $\eta$ is the nonce of the deploying transaction. $\eta$ is always equal to the number of transactions sent from the deploying EOA. As a result, we can deterministically calculate the address of a future smart contract that will be deployed by a certain user.

\noindent\textbf{EVM Function Selector.} 
In EVM, when a smart contract function is called by an EOA or another smart contract, the calling function is identified by its \emph{selector} $S_f$ as follows:
$$S_f = P_{32}(H_{k}(\overbrace{\text{``}f(\alpha_1,...,\alpha_n)\text{''}}^{\text{\normalsize \shortstack{function header string}}})),$$
where $P_{32}$ is a 32-bit prefix, $H_{k}$ is the \mbox{Keccak256} hash function, $f$ is the function name, and $\alpha_1,...,\alpha_n$ is the list of argument types ($0 \le n \le 16$). For example, the selector of the function $foo$ with a single 256-bit unsigned integer argument is $P_{32}(H_{k}(``foo(uint256)"))=\mathrm{0x2fbebd38}$.

%% file: threatmodel.tex
\section{Threat Model}\label{sec:threat-model}
In this section, we give a general overview of social engineering attacks in Ethereum smart contracts by identifying their participants, vectors, goals, and outcomes.

\subsection{Actors}

Most known attacks in Ethereum smart contracts involve a hacker exploiting a smart contract vulnerability~\cite{zhou2020ever,antonopoulos2018mastering}. In social engineering attacks, however, a reverse configuration takes place: the owner of the malicious smart contract is the attacker, and the victim of the smart contract is a person or organization who engages with this smart contract.

\subsection{Social Engineering Attack Vectors}

Here, we expose a number of social engineering attack vectors that are likely to be exploited. Essentially, all these vectors are misconceptions (false assumptions) about properties or behaviors of the Ethereum platform. We subdivide these misconceptions into two major categories: 1) misconceptions about Ethereum addresses, and 2) misconceptions related to strings and characters in EVM and Solidity.

\noindent\textbf{Misconceptions About Addresses.} An Ethereum public address is a 160-bit number using a 40-digit hexadecimal representation. Our analysis reveals that the following four false assumptions about Ethereum public addresses can be exploited in social engineering attacks. 

\begin{itemize}
    \item \msI: \textit{Slight modification of an address (e.g., substitution of a single digit) is useless for an attacker because no one knows the private key associated with the modified address.} In Section~\ref{subsec:attackI}, we demonstrate that the knowledge of the private key for an address is not always required for a successful social engineering attack.
    
    \item \msII: \textit{EIP-55 checksums deliver a reliable protection against address falsification.} In Sections~\ref{subsec:attackI} and \ref{subsec:attackIII}, we show that EIP-55 falsification is possible using a brute-force attack on a retail laptop or desktop computer.
    
    \item \msIII: \textit{An Ethereum address is associated either with an EOA, or a smart contract, and does not change its status.} In Section~\ref{subsec:attackII}, we demonstrate that an EOA can mutate into a smart contract and vice versa.
    
    \item \msIV: \textit{All Ethereum accounts are equally secure as long as their private keys are random and secret.} In Section~\ref{subsec:attackIII}, we show that a small portion of Ethereum accounts have a special property, making them more vulnerable to a specific social engineering attack.
\end{itemize}

\noindent\textbf{Homograph Backdoors in Solidity.} Falsification of typographic symbols, known as \emph{homograph} or \emph{Unicode} attacks, have been used in phishing scams ~\cite{fu2006methodology,holgers2006cutting,liu2007fighting}. These attacks mostly falsify domain names, and \emph{to the best of our knowledge, there are no recorded homograph attacks carried inside a source code of a program.} Surprisingly, our analysis of Solidity reveals the following three misconceptions that open dangerous backdoors to homograph attacks in Ethereum smart contracts.

\begin{itemize}
    \item \msV: \textit{Since the string returned by the ERC-20 \ttt{symbol()} function is optional and informational by design, it does not pose any danger.} In Section~\ref{subsec:attackIV}, we show that by falsifying the symbol of an ERC-20 token, an attacker can perform a social engineering attack.
    
    \item \msVI: \textit{Two identical arguments of \ttt{call()} or \ttt{delegatecall()} always result in the same 32-bit function selector.} In Sections~\ref{subsec:attackV} and \ref{subsec:attackVI}, we demonstrate that two identical arguments are capable of producing different function selectors, which leads to the execution of an unexpected function or transaction reversion due to the absence of a referenced function.
    
    \item \msVII: \textit{Function selector collision prevention by Solidity compiler eliminates falsification of smart contract functions.} In smart contracts, two functions with colliding selectors cannot coexist in one contract. In Sections~\ref{subsec:attackV} and \ref{subsec:attackVI}, we show that it is possible to mine names of two functions with \emph{visually identical} arguments of \ttt{call()} or \ttt{delegatecall()} routines that generate different selectors, thereby allowing these two functions to coexist in the contract. Consequently, unbeknownst to the transaction sender, a non-existent function might be called, resulting in transaction reversal; or a wrong function might be called, leading to unexpected code execution.
\end{itemize}

\subsection{Attack Goals and Outcomes}

Although some Ethereum attackers may pursue vandalism as the primary goal (e.g., via "funds freeze"), in this work, we assume that the ultimate objective of the attacker is \emph{to steal funds from victims}. All social engineering attacks covered in this study are based on the premise that \emph{the attacker is the owner or privileged user of the smart contract}\footnote{In Ethereum, the implementation of smart contract ownership is the developer's responsibility. Zhou et al.~\cite{zhou2020ever} report more than 2 million contracts with ownership implemented using the OpenZeppelin \texttt{Ownable} abstract class and \texttt{onlyOwner} modifier.}, which creates a broad range of possibilities for stealing funds. For example, many contracts implement the \ttt{selfdestruct} procedure, which allows the owner to appropriate the entire balance of the contract by submitting a single transaction. 

Moreover, as of early December 2020, Etherscan reports more than 342,000 ERC-20 smart contracts, which have a variety of operations with tokenized funds, such as minting, burning, approved transfer, etc. For example, in Tether USD stablecoin token, which is worth over \$19 billion, the owner can call the \ttt{deprecate} function of the contract, effectively replacing the functionality of the smart contract into any arbitrary code. Subsequently, it would take only a few minutes for the contract owner to steal all the tokens and exchange them into Ether, at which point no existing defense can revert the theft of funds. Essentially, when the attacker is the owner of the smart contract, it is unnecessary to implement the malicious transfer of funds within the call stack of the transaction submitted by the victim. Instead, the attacker may prefer to accrue a sufficient sum by blocking fund withdrawals, and acquire the entire balance afterwards. Such an approach makes the malicious patterns more stealthy than an immediate transfer of stolen funds.

%% file: attacks.tex
\vspace{-5pt}
\section{Social Engineering Attacks}\label{sec:se-attacks}

In this section, we introduce six Ethereum social engineering attacks grouped into two classes, as shown in Table~\ref{table:attacks}. The \emph{Address Manipulation class} allows attackers to strategically exploit Ethereum public addresses, which empowers attacks \attackI, \attackII, and \attackIII. The \emph{Homograph class}, which takes advantage of the fact that many fonts have identically looking symbols with different codes, includes attacks \attackIV, \attackV, and \attackVI. The implementations of all the six attacks are available at \url{https://nick-ivanov.github.io/se-info/}.

\begin{table*}[t]
\centering
\caption{Social engineering attacks in Ethereum smart contracts.}
\begin{tabular}{|c|l|c|}
\hline

\multirow{2}{*}{\textbf{\small Attack Class}} & \multicolumn{1}{|c|}{\multirow{2}{*}{\textbf{\small Social Engineering Attacks and Brief Descriptions}}} & \textbf{\small Misconceptions} \\
&  & \textbf{\small Exploited} \\
\hline \hline
       
\multirow{3}{*}{\shortstack{Address\\Manipulation}} & \emph{\attackI: Replace EOA with a non-payable contract address to incur transfer failure and revert transaction} & \msI, \msII \\

\cline{2-3}

& \emph{\attackII: Pre-calculate a future contract address and replace EOA with a non-payable contract at this address} & \msIII  \\

\cline{2-3}

& \emph{\attackIII: Exploit EVM's EIP-55 checksum insensitivity in address comparison} & \msIV \\

\hline

\multirow{3}{*}{Homograph} & \emph{\attackIV: Use dynamically-injected homograph string in a branching condition} & \msV \\

\cline{2-3}  

& \emph{\attackV: Replace inter-contract call (ICC) header with identically looking one to call a non-existing function} & \msVI, \msVII \\

\cline{2-3}

& \emph{\attackVI: Suppress EVM exception by mining a function that matches a tampered ICC header} & \msVI, \msVII  \\

\hline
                                    
\end{tabular}
\label{table:attacks}
\end{table*}

\noindent \textbf{Base Token.}
We demonstrate all the six attacks by altering the implementation of the smart contract called \emph{Base Token} (see Fig.~\ref{fig:basetoken}). This contract is an Ether-collateralized ERC-20 token, which means that the supply of tokens in the contract is backed by its Ether balance, allowing users to swap (i.e., buy and sell) the tokens using Ether. We implement Base Token using the OpenZeppelin ERC-20 prototype with two additional methods:

\begin{itemize}
    \item \ttt{buyToken} method deposits Ether in the smart contract and mints (issues) tokens corresponding to the deposited amount;
    \item \ttt{sellToken} method burns (destroys) user tokens and transfers the corresponding amount of Ether to the caller.
\end{itemize}

\begin{figure}[t]
    \centering
\begin{lstlisting}[language=Solidity]
contract BaseToken is Context, ERC20, ERC20Detailed {
  uint256 tokenPrice = 100 wei;
  constructor () public payable ERC20Detailed(     "BaseToken", "BT", 18) {
    _mint(_msgSender(), SafeMath.div(msg.value, tokenPrice));
  }
  function buyTokens() public payable {
    _mint(_msgSender(), SafeMath.div(msg.value, tokenPrice));
  }
  function sellTokens(uint256 amount) public {
    _burn(_msgSender(), amount);
    address(msg.sender).transfer(SafeMath.mul(amount, tokenPrice));
  }
}
\end{lstlisting}
    \caption{Implementation of the Base Token, which is used to demonstrate the six social engineering attacks.}
    \label{fig:basetoken}
\end{figure}

\subsection{Address Manipulation}

\emph{Address Manipulation} attacks exploit cognitive biases and misconceptions about equality, format, referenced objects, derivation methods, and other properties of Ethereum public addresses.
In this section, we propose three social engineering attacks: \attackI, \attackII, and \attackIII.

\subsubsection{Attack \attackI}\label{subsec:attackI}

\begin{figure}[t]
    \centering
    \includegraphics[width=\linewidth]{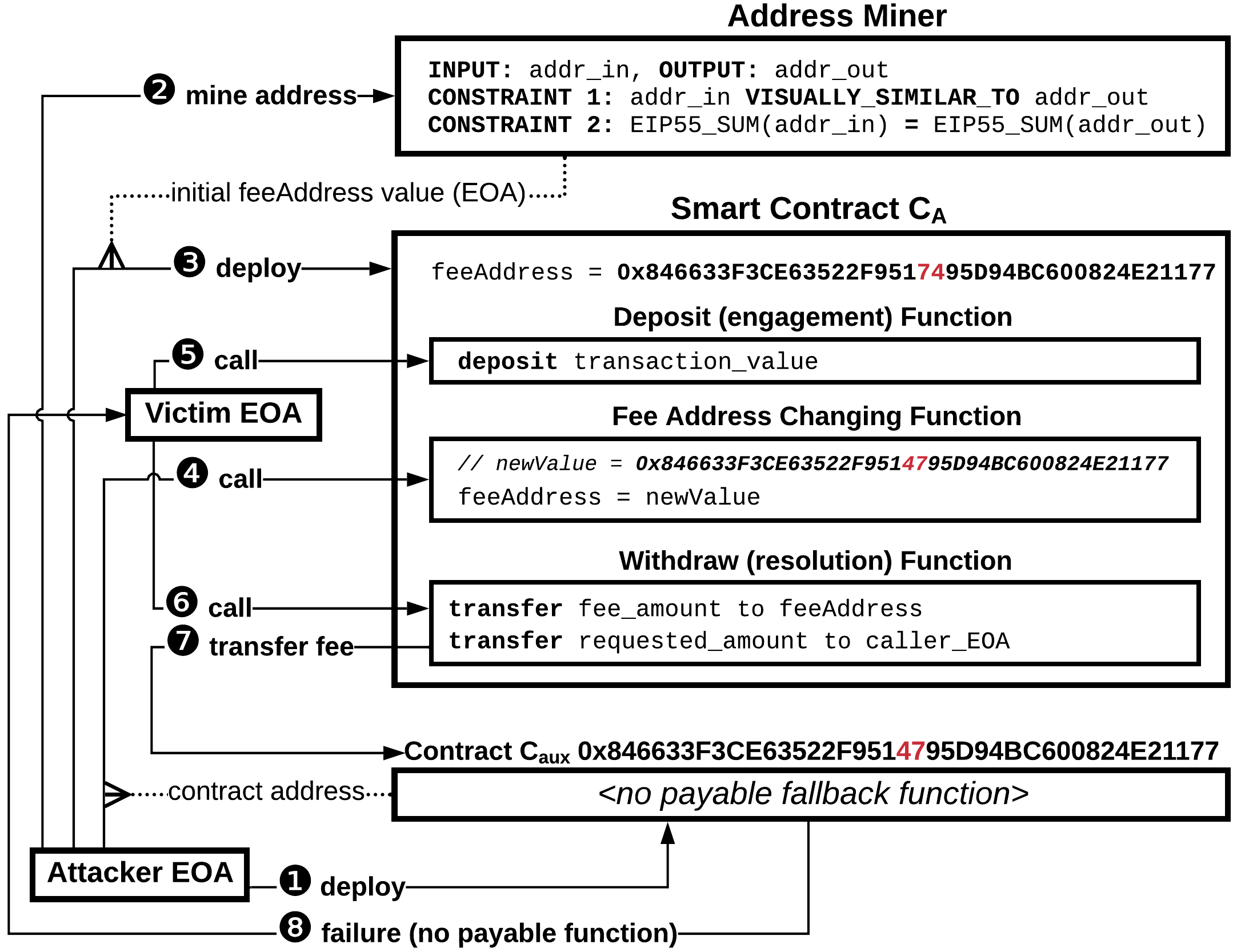}
    \caption{Attack \attackI workflow.}
    \label{fig:a1-workflow}
\end{figure}

\emph{This attack covertly substitutes an EOA address into a similar smart contract address that allows the attacker to block funds withdrawal and subsequently acquire them.} In \attackI attack, the attacker deploys a smart contract with two sequential Ether transfers within the call stack of one transaction. The first transfer \emph{looks like} a fee collection, while the second transfer is a fund transfer to the user. The attacker deceives a victim to believe that the first transfer goes to an EOA, whereas the real destination is a smart contract without a payable fallback function. Therefore, the transfer fails, and the funds (deposited by the users earlier) remain in the malicious contract, which are available for the attacker for subsequent withdrawal through contract self-destruction, deprecation, or similar mechanism.

Essentially, the attacker exploits the fact that almost any unused sequence of 40 hexadecimal digits is a valid EOA address, even if its corresponding private key is unknown. Particularly, if a few symbols in an address are replaced or swapped, the resulting address will still be a valid Ethereum EOA, which can accept incoming Ether transfers. In \attackI attack, as shown in Fig.~\ref{fig:a1-workflow}, the adversary deploys a malicious smart contract $C_A$. The variable \ttt{feeAddress} in this contract is initiated with an EOA address $A_1$. Also, each fund transfer to the user is preceded by another transfer of a small fee to the address stored in \ttt{feeAddress}. This creates a perfect illusion that the smart contract was deployed to profit from service fees. However, the real purpose of the contract is to lure the user to make a deposit and block any attempt to withdraw the funds.

To achieve that, we introduce another public address $A_2$, derived from address $A_1$ by either changing one symbol or swapping neighboring symbols to make two addresses \emph{visually similar}. The manipulated address must maintain a valid checksum \textit{that collides with the checksum of the original address}, reassuring the user that the address is the one seen in the constructor. We find that mining such an address pair takes only a few seconds\footnote{Our address miner is available at \url{https://github.com/nick-ivanov/se-tools}}, and thus demonstrate the incorrectness of \msII. Address $A_2$ belongs to a pre-existing smart contract $C_{aux}$, which does not have a payable fallback function. The attacker sets the value of \ttt{feeAddress} into $A_2$. Due to the addresses' visual similarity, the user deposits funds with the assumption that the fees go to $A_1$. However, the withdrawal fails due to an attempt to send fees to an unpayable smart contract. For further deception, the attacker can generate a history of successful fee transfers from the smart contract to address $A_1$, deceiving  the users into believing that the smart contract is actively receiving successful fee payments. This deepens the users' confirmation bias that complies with the attacker's deceptive narrative.

The attack workflow in Fig.~\ref{fig:a1-workflow} includes four layers of deception that give the victim several clues aligned with the same narrative (i.e.,  the contract is a fair for-profit scheme), thereby exploiting the victim's confirmation bias. The first layer of deception is that the smart contract does not reveal its deceptive nature during a test deployment --- if a user compiles and deploys this smart contract for testing, the scheme will support the deceptive narrative because the test deployment cannot predict that the owner would change the value of \ttt{feeAddress} into the address of a non-payable smart contract. The second layer of deception comes from the deployment-time initialization of the \ttt{feeAddress} variable: by examining this address, the victim finds a history of fair transactions. The third layer of deception is delivered through keeping the \ttt{feeAddress} variable private, which prevents the victim from easy retrieval of its current value, as it requires a laborious effort of parsing binary transaction data. The fourth layer of deception targets a user who manages to retrieve the current value of \ttt{feeAddress}. Since this value is visually similar to the initialization address, the victim is likely to conclude that the original address is in use.

\subsubsection{Attack \attackII}\label{subsec:attackII}

\emph{This attack intercepts a client deposit event and immediately deploys an auxiliary malicious smart contract at an EOA address for stealing funds accrued via blocked withdrawals.} The key idea is to mislead the user by runtime replacement of \emph{what an address points to}. The attack utilizes a more sophisticated method that dynamically changes the object referenced by an address. Here, we discover a peculiar combination of two facts about Ethereum that lead to the incorrectness of \msIII: a) the address of a future, not yet deployed, smart contract is predictable; b) prior to deployment, the address of the future smart contract has the status of a legitimate EOA. Recall from Section~\ref{sec:background} that a smart contract address is generated from the address of the deploying EOA and the transaction tally in this EOA.

\begin{figure*}[t]
    \centering
    \includegraphics[width=6.75in]{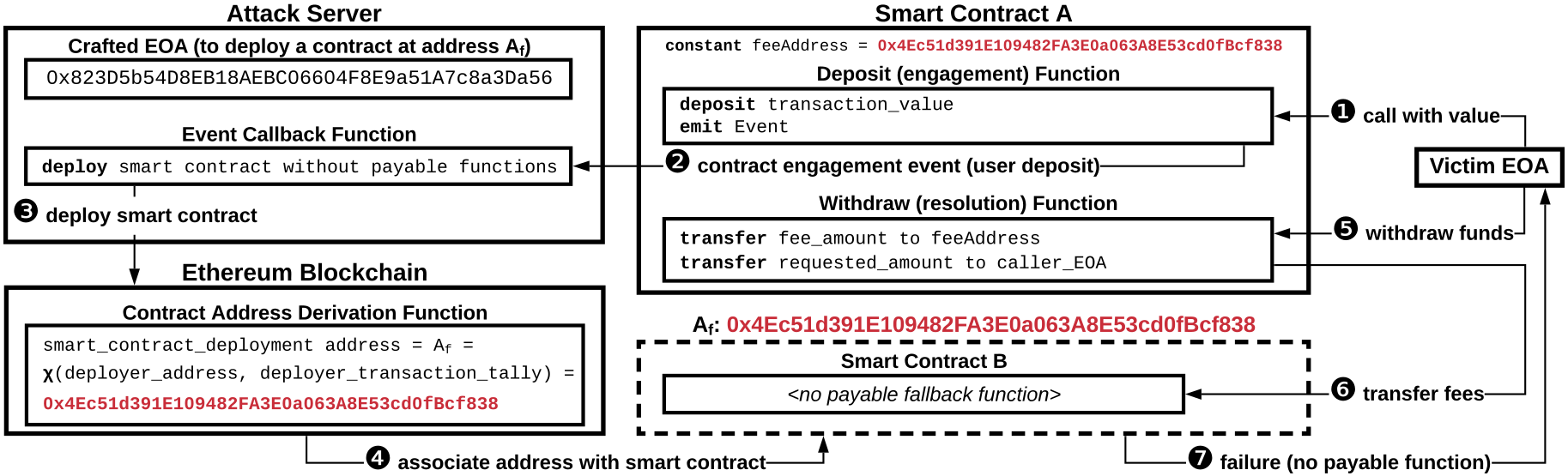}
    \caption{Attack \attackII workflow.}
    \label{fig:eavesdropper-bev}
\end{figure*}

Fig.~\ref{fig:eavesdropper-bev} illustrates the workflow of attack \attackII. Smart contract A is disguised as a fair for-profit scheme, in which the owner charges fees per fund withdrawal. The fee recipient address is hard-coded in the smart contract and set as a constant, which 
fuels the confirmation bias supporting the notion of permanence of this address. For normal operation, this address should accept incoming funds, which means that it should either be an EOA or a smart contract with a payable fallback function. When the user makes a deposit, an event is emitted, which is intercepted by a server belonging to the attacker (the owner of smart contract A). Upon the detection of the event, the attacker deploys smart contract B at the address $A_f$. The fee collector address $A_f$ is crafted in a way that the attacker knows the corresponding private key of the account $A_d$, based on which the contract B is deployed, i.e., $A_f = \chi(A_d,\eta)$ (see Section~\ref{sec:background}). The fee transfer to address $A_f$ now fails because smart contract B has no payable fallback function.  As a result, the previously deposited funds remain in the contract for subsequent acquisition by the attacker.

\subsubsection{Attack \attackIII}\label{subsec:attackIII}

\emph{The attack leverages the overlap between lower-case and mixed-case EIP-55 addresses to misguide users into locking their funds in the smart contract for subsequent acquisition thereof by the attacker.} In attack \attackIII, the attacker provides the user with a personal smart contract and a \emph{seemingly random} test Ethereum accounts. When a smart contract has hard-coded addresses or other account-specific values, it is a common practice to provide users with test accounts to demonstrate the functionality of a smart contract~\cite{antonopoulos2018mastering}. Since all accounts are assumed to have the same set of properties, the user believes that any account will have the same behavior as the test accounts, which we found not to be always true. Essentially, attack \attackIII exploits \msIV, i.e., the belief that the secrecy  of the private key solely determines the security of an Ethereum account. The key to this attack is the generation of accounts with all lowercase EIP-55 checksums. We verify that the probability of generating an EIP-55 address with lowercase checksums is about $0.0246\%$ using a random guessing approach. Please refer to Appendix B for examples of such addresses.

One-time-password validation is a common supplemental authorization technique in smart contracts\footnote{Sample password-based authorization can be found in these contracts:\mbox{~}\texttt{0x0f82C7EAb8F7efB577A2DE9d2B7e1Da1d0b6870e},\mbox{~}and \texttt{0x13407d93F343148bf03eaCf482441dD526cD7EbD}.}. The smart contract owner can generate an \emph{authentication hash} of the user address and the corresponding user password, and store this hash in the smart contract.  In this attack, the adversary creates such a password validation routine in the smart contract, and offers the user several test accounts for verification of functionality. However, the test set consists of only deliberately mined accounts with \emph{all lowercase EIP-55 checksums}. In this smart contract, the fund transfer function is preceded by a password validation, which invokes an address conversion function that translates the address of the transaction sender into an all-lowercase string (e.g., \ttt{strAddrHash} in Fig.~\ref{fig:blackbox}). Using the test accounts, the smart contract works as expected. After the testing, the user creates a production authentication hash by concatenating his/her public address (copied from the wallet) and a secret password. This production account cannot be tested to avoid revealing the password through the open network of the public blockchain. Unexpectedly, an attempt to withdraw the funds will fail due to a failure in password validation caused by the disparity in the address capitalization.

Fig.~\ref{fig:blackbox} demonstrates an example of attack \attackIII. The \ttt{authHash} constant variable stores the Keccak256 digest of the user address $0xe6c700856796524501438d7197497c14bceac297$ concatenated with the password \ttt{ASIACCS2021}. The attacker offers the user the private keys of several test accounts,  whose public addresses' EIP-55 checksums are all lowercase. These test accounts work as expected. But when the users initiate transactions with their real addresses, the password validation fails, since \ttt{authHash} incorporates the address with checksums in mixed-case letters, while \ttt{strAddrHash} generates the hash using the same address with all lowercase checksums. This failed validation prevents the selling of tokens by the user. This attack demonstrates that some accounts can be more vulnerable than others, effectively defying misconception \msIV.

\begin{figure}[t]
    \centering
\begin{lstlisting}[language=Solidity]
bytes32 constant authHash =
0x8e69860da968defb8d06a7e565e5d76e3e878a01473   a0cb191a0eda120323ca5
function strAddrHash(address _addr,
string memory _pass) private pure returns (bytes32) {
  return keccak256(abi.encodePacked(addr2Str(_addr), _pass));
}
function sellTokens(uint256 amount, string memory password) public {
  if(strAddrHash(msg.sender, password) == authHash) {
    _burn(_msgSender(), amount);
    address(msg.sender).transfer(SafeMath.mul(amount, tokenPrice));
  }
}
\end{lstlisting}
    \caption{Code snippet from function \ttt{sellTokens} in \attackIII attack.}
    \label{fig:blackbox}
\end{figure}

\subsection{Homograph Visual Cognitive Deception}

The homograph attacks in smart contracts are enabled by the existence of symbols that look identical or very similar, whereas most text editors (except hex viewers) are unable to reveal the difference. We surveyed security experts from seven smart contract auditing firms (listed in Section~\ref{sec:firms}) about the usage frequency of hex viewers in their auditing process. The survey results show that only 1 out of 7 companies uses hex viewers \emph{usually}, 2 of them use hex viewers \emph{sometimes}, while the rest \emph{never} or \emph{rarely} use them. Here, we define two words or letters that contain identically looking symbols with different codes as a pair of \emph{homograph twins}. The Homograph class of social engineering attacks leverages the fact that: although Solidity prohibits Unicode symbols in the names of functions and variables, it allows these symbols to appear in string literals that determine branching and inter-contract calls. In this section, we introduce three Homograph attacks: \attackIV, \attackV, and \attackVI.

\subsubsection{Attack \attackIV}\label{subsec:attackIV}
\emph{The attack leverages homograph twins in a string matching pattern to craft a malicious smart contract}.
Specifically, the attacker crafts a smart contract in which a homograph string is used in a branching condition, which leads to unexpected code execution.

Fig.~\ref{fig:badchoice20} demonstrates attack \attackIV, with the attack code embedded in the \ttt{sellTokens()} function.
The \ttt{stringsEqual()} function performs a string matching by comparing the hashes of two strings\footnote{Solidity does not have any embedded or library string matching function. As Keccak256 digest is an EVM opcode function with relatively low gas cost, comparing string hashes is de-facto the standard string comparison approach.}. The literal \ttt{BT} is made of two ASCII characters, but the \ttt{symbol()} return value, although visually identical to literal \ttt{BT}, has the symbol \ttt{T} substituted with its homograph twin from the Cyrillic symbol set. Since the value of \ttt{symbol()} is mutable, the smart contract does not contain any explicitly malicious code, however, it \emph{turns} malicious when the token symbol value is changed. As a result, the branching condition turns false, and the sell of tokens never occurs, which proves the importance of the token symbol, and thus refuting misconception \msV.

\begin{figure}[t]
    \centering
\begin{lstlisting}[language=Solidity]
if(stringsEqual(symbol(), "BT")) {
  _burn(_msgSender(), amount);
  address(msg.sender).transfer(SafeMath.mul(amount, tokenPrice));
}
\end{lstlisting}
    \caption{A snippet of the \ttt{sellTokens} function in \attackIV attack.}
    \label{fig:badchoice20}
\end{figure}

\subsubsection{Attack \attackV}\label{subsec:attackV}

\emph{This attack replaces the header of a function with its homograph twin to cause unexpected inter-contract call failures.}
\emph{Code reuse} has been one of the best practices of smart contract development, allowing to reduce implementation time and frequency of programming errors. Code reuse can be either static or dynamic. A typical example of static code reuse is inheriting classes from the OpenZeppelin Contracts library. EVM also supports dynamic code reuse, in which one smart contract calls functions of another contract deployed on the same blockchain. Dynamic code reuse reduces the utilization of blockchain storage and achieves native \emph{inter-contract communication (ICC)}. It is known that if a function is specified incorrectly in an ICC call, the fallback function\footnote{In Ethereum smart contracts, the fallback function is an optional nameless function  designed to be a default interface of a smart contract.} of the smart contract will be invoked instead~\cite{atzei2017survey}. However, if the fallback function is absent, the call to a non-existent function triggers an EVM exception with subsequent transaction reversal, which is utilized by attack \attackV via falsification of a function ICC selector.

\begin{figure}[t]
    \centering
    \includegraphics[width=3.3in]{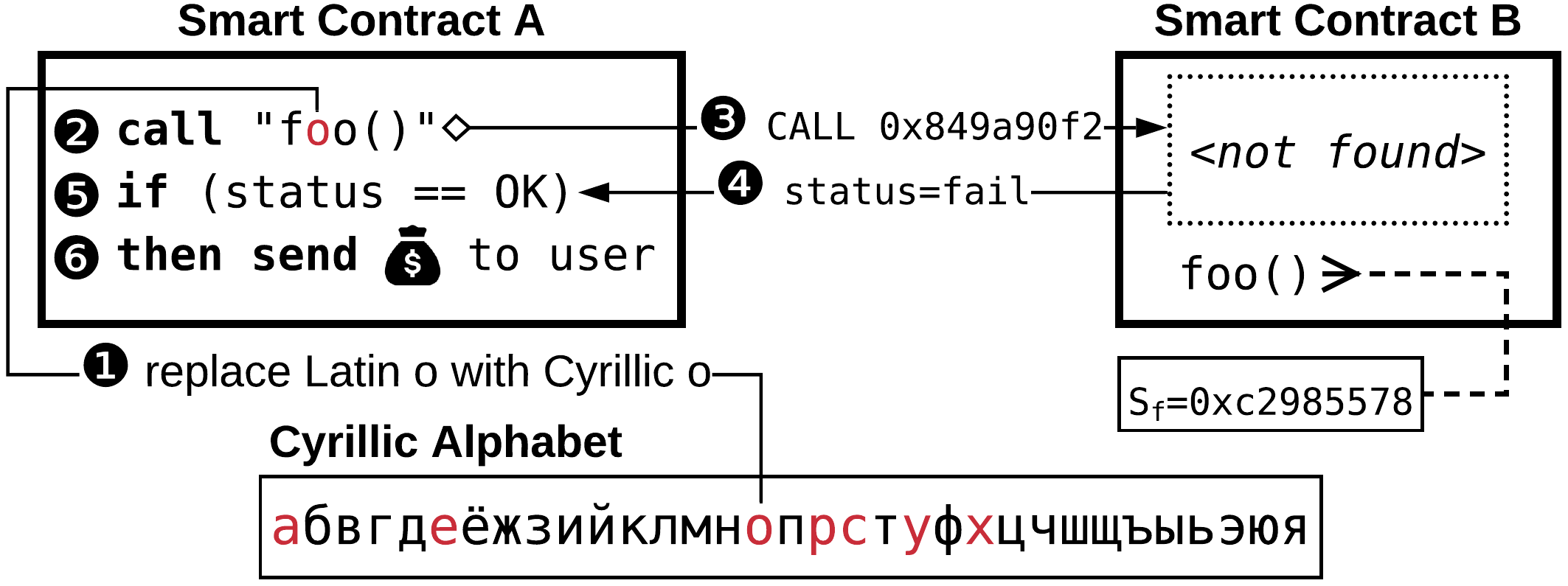}
    \caption{Attack \attackV workflow.
    }
    \label{fig:callofthevoid-bev}
\end{figure}

Fig.~\ref{fig:callofthevoid-bev} demonstrates the general idea of attack \attackV. During an ICC call, when an expected function in the destination smart contract is not found, and with no fallback routine implemented, the call will unexpectedly fail, and the transfer of funds to the client will not be executed. The proposed \attackV attack substitutes one or several letters in the function header string with homograph twins, and as a result, the generated function selector will not match any existing function, leading to the ICC call failure.

\begin{figure}[t]
    \centering
\begin{lstlisting}[language=Solidity]
bytes memory payload = abi.encodeWithSignature
("log(address)", msg.sender);
bool success = address(helperAddress).call(payload);
if(success) {
  _burn(_msgSender(), amount);
  address(msg.sender).transfer(SafeMath.mul(amount, tokenPrice));
}
\end{lstlisting}
    \caption{Code snippet from function \ttt{sellTokens} in \attackV.}
    \label{fig:callofthevoid20}
\end{figure}

Fig.~\ref{fig:callofthevoid20} shows the \ttt{sellTokens} function of \attackV attack. We create and deploy an additional smart contract called \ttt{Helper} (see Fig.~\ref{fig:callofthevoid-helper}), whose address is hard-coded in the \ttt{BaseToken} contract. The \ttt{Helper} smart contract has a \ttt{log} function for event logging. However, the string ``\ttt{log(address)}'' contains letters substituted with their homograph twins, and therefore the ICC call fails. Thus, the subsequent fund transfer to the caller never happens. This example demonstrates that visually identical arguments of \ttt{call()} and \ttt{delegatecall()} routines can indeed produce different selectors, proving the incorrectness of \msVI.

\begin{figure}[t]
    \centering
\begin{lstlisting}[language=Solidity]
mapping(address => uint256) private lastSell;
function log(address a) public {
  require(msg.sender ==
    0x0EFb5DE6AddAdDE835CEaadaAB1992590d7588F5);
  lastSell[a] = block.number;
}
\end{lstlisting}
    \caption{A code snippet of the \ttt{Helper} contract used in \attackV.}
    \label{fig:callofthevoid-helper}
\end{figure}

\subsubsection{Attack \attackVI}\label{subsec:attackVI}
The previous attack has one major weakness: although nothing in the code looks suspicious, the status check of the ICC call may prompt a cautious user to set up a test deployment to check whether the call succeeds or not. Our next attack provides a deceptive technique to pass such a test. \emph{Attack \attackVI leverages potential collision cases of Ethereum function selectors, whose length is only 32 bits, to ensure a successful status from a deceptive ICC call.} Assuming a uniform distribution of function selectors, the probability of collision with another function (i.e., two functions have the same selector) is approximately $2.33 \cdot 10^{-10}$. We run an experiment to show that it only takes a few hours on average for an office computer to find a collision\footnote{Generally, the larger the number of symbols available for homograph substitution in the function header, the less time it takes to mine a collision.}. In attack \attackVI, the attacker crafts a function whose selector collides with the selector of the homograph twin of the expected function. Since the called function actually exists, the transaction succeeds, which further fuels the confirmation bias of the victim supporting the deceptive narrative crafted by the attacker.

\begin{figure}[t]
    \centering
    \includegraphics[height=2.5in]{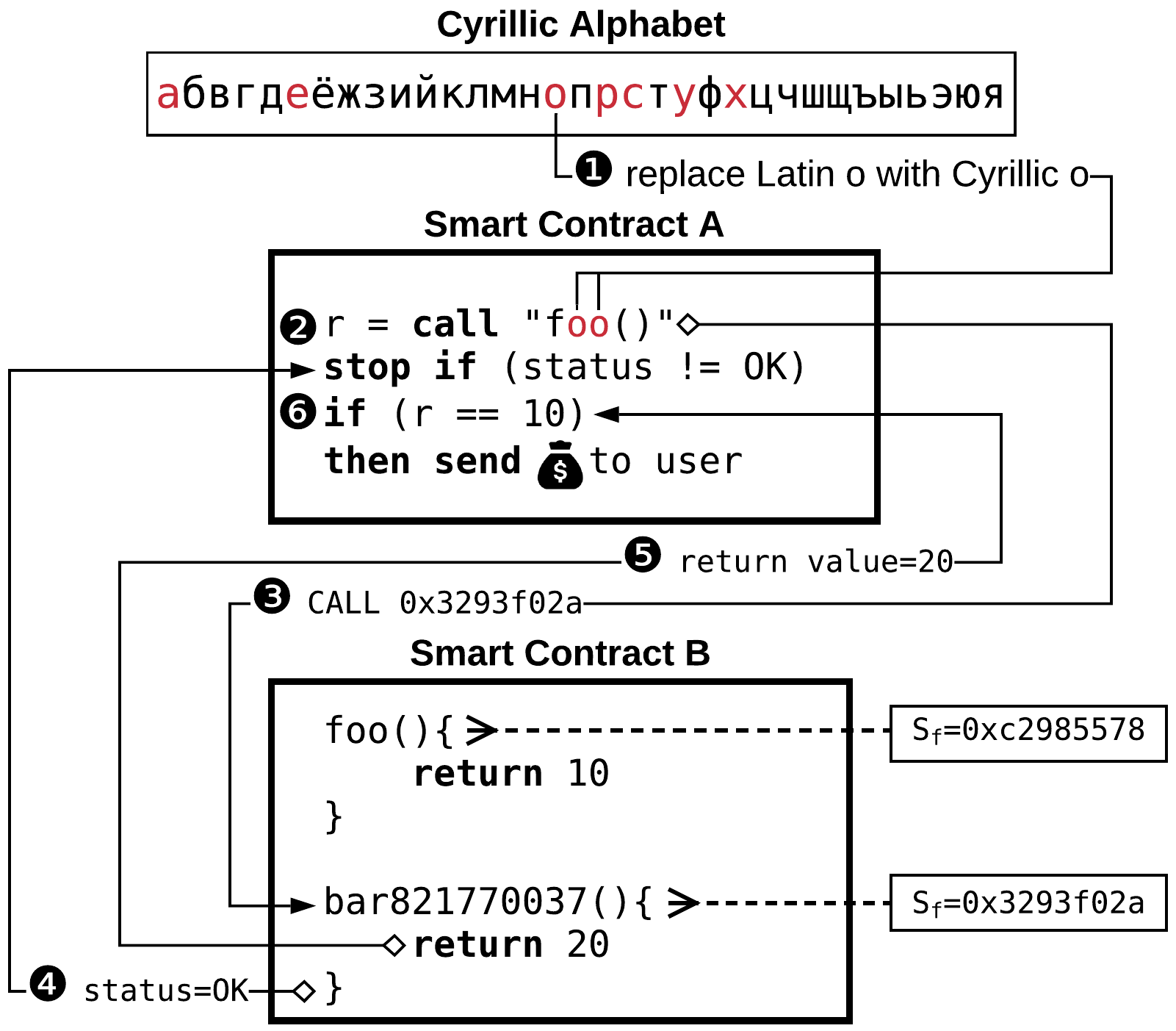}
    \caption{Workflow in the \attackVI attack.
    }
    \label{fig:namecalling-bev}
\end{figure}

The Solidity compiler will terminate with an error if it encounters two functions with the same selectors in one smart contract. \attackVI attack avoids this issue by replacing a function header with its homograph twin. In the workflow of the attack, presented in Fig.~\ref{fig:namecalling-bev}, smart contract A implements a call to a function in smart contract B. When B is compiled, the string header of the function \ttt{foo} will be translated into the 32-bit selector $0xc2985578$. However, if we substitute both the letters ``o'' in the string ``\ttt{foo()}" with their homograph twins, the compiler will translate the modified header into the selector $0x3293f02a$. Now, the attacker uses a collision search algorithm to mine the function name \ttt{bar821770037}, whose selector is also $0x3293f02a$. As a result, \ttt{foo} and \ttt{bar821770037} can coexist in contract B, despite the fact that they both have visually identical argument of \ttt{delegatecall}, i.e., \ttt{"foo()"} (see step \ding{183} in Fig.~\ref{fig:namecalling-bev}), effectively refuting \msVII. After the homograph substitution, unbeknownst to the user, \ttt{bar821770037} will be called instead of \ttt{foo}, which will return a successful status but break the anticipated code logic in contract A.

\begin{figure}[t]
    \centering
\begin{lstlisting}[language=Solidity]
bytes memory payload = abi.encodeWithSignature
  ("accountRegistered(address)",msg.sender);
(bool success, bytes memory result) = address(helperAddress).delegatecall(payload);
require(success);
if(abi.decode(result, (bool)) == true) {
  _burn(_msgSender(), amount);
  address(msg.sender).transfer(SafeMath.mul(amount, tokenPrice));
}
\end{lstlisting}
    \caption{Code snippet from function \ttt{sellTokens} in \attackVI.}
    \label{fig:namecalling}
\end{figure}

Figs.~\ref{fig:namecalling} and~\ref{fig:namecalling-helper} demonstrate an example of the \attackVI attack. The \ttt{Helper} smart contract includes two functions, \ttt{accountRegistered} and \ttt{afterBlock29410106}. Since block number checks are common in Ethereum smart contracts\footnote{For example, contract \texttt{0xb68c88283b558cdc38c75c07bbc0d6921ef40fc7} uses a block number check to determine the contract initialization deadline.}, the presence of an auxiliary function with this name is unlikely to raise any suspicion. The string ``\ttt{accountRegistered(address)}'' (Fig.~\ref{fig:namecalling}) contains Cyrillic letters (letters 1, 2, 3, and 16 are replaced). We use a brute-force algorithm to mine the name \ttt{afterBlock29410106}, whose function selector collides with a homograph twin of \ttt{``accountRegistered(address)''}. Surprisingly, we discover that the functions \ttt{afterBlock29410106} and \ttt{accountRegistered} can accept arguments of different types: the call will still succeed regardless of the argument types, as long as the number of arguments in the two functions is consistent. \emph{This undocumented behavior of EVM adds an additional layer of disguise to the attack.} In the end, \ttt{afterBlock29410106} is called instead of the expected function \ttt{accountRegistered}. Unlike in \attackV attack, the \ttt{success} variable is now \emph{true}. However, the user's fund transfer does not happen despite the successful return status, as the function's return value is not as expected.

\begin{figure}[t]
    \centering
\begin{lstlisting}[language=Solidity]
function afterBlock29410106(bool deadlineCheck)
  public view returns (bool) {
  if(block.number > 29410106 && deadlineCheck) {
    return true;
  }
  return false;
}
function accountRegistered(address a) public pure returns (bool) {
  return a==mainAccount || a==backupAccount;
}
\end{lstlisting}
    \caption{A snippet of the Helper contract used in \attackVI attack.}
    \label{fig:namecalling-helper}
\end{figure}

%% file: exploitation.tex
\section{Case Study of Real-world Smart Contracts}\label{section:exploitation}

\begin{table}[t]
    \caption{Five popular tokens that we succeed in integrating social engineering attack patterns.}
    \label{tab:exploitation-mini}
    
    \centering
    \begin{tabular}{|c|c|c|}
    \hline
    \textbf{Smart}  & \textbf{Market}\textsuperscript{\dag} & \textbf{Integrated}\\
    \textbf{Contract} 
     & \textbf{Cap.} \textit{($\times$\$1 billion)} & \textbf{Attack Pattern} \\
    \hline
    \hline
     Tether USD (USDT) & 19.76 & \attackIV \\     
     \hline

     Binance (BNB) &    4.6 & \attackV \\
     \hline

     ChainLink (LINK) & 3.94 & \attackI \\
     \hline

     Bitfinex (LEO) &  1.32 & \attackVI \\
     \hline
     
     CryptoKitties (CK) & --- & \attackI + \attackII \\
     \hline
     
     \multicolumn{3}{l}{\textsuperscript{\dag} Approximate rounded averages as of early December 2020.
     } \\
    \end{tabular}
\end{table}

One of the most important questions of this paper is whether the six social engineering attacks can be used in real-world smart contracts. To answer this question, we choose source codes of five smart contracts that meet the following criteria: a) they represent a popular use case of a smart contract; b) they have thousands of active users; c) they have high market capitalization (i.e., the users entrust them their funds); d) the contracts implement one of the standard use cases from the OpenZeppelin Contract library. Then, we slightly modify the source codes of these contracts to integrate the social engineering attacks into them without altering any functionality or incorporating any unsafe practices or known vulnerabilities. This way we demonstrate that popular trusted smart contracts are capable of delivering the social engineering attacks.

After integrating the attack patterns into the source codes of the five contracts, we deploy the contracts on Ropsten testnet and validate their expected functionalities. Then, we simulate the production deployment of the contracts, and demonstrate that some transactions that worked during the testing will fail due to activation of the attack functionality (e.g., deployment of a contract at EOA address in attack \attackII). For each case, we make sure that:
a) the attacks remain dormant during the test stage and activate only on a production deployment;
b) the attacks visually conceal themselves from the auditor; and c) each attack has a rational disguise (e.g., pretend to profit from charging service fees). Table~\ref{tab:exploitation-mini} summarizes the five smart contracts and attack patterns integrated in them.  Appendix C provides more details of evasive testing and exploitation demonstration for each of the five cases. The video demonstrations of all the five cases are available at \emph{\url{https://nick-ivanov.github.io/se-info/}}. The source code files of the entire smart contract set are available at \emph{\url{https://github.com/nick-ivanov/social-engineering-big5}}.

\noindent\textbf{Production Deployment Simulation.}
Our manual analysis of the source codes of popular contracts reveals that most of them use
the OpenZeppelin Contracts templates with some custom additions. In our case study, we demonstrate the feasibility of an attack code integration into an existing token without breaking the security patterns and functionality delivered by the OpenZeppelin Contracts library.
The manipulated token can be advertised as a new cryptocurrency with additional features, such as special VIP privileges for early adopters. For ethics concerns, we perform both testing and production deployment simulation using the Ropsten testnet, whose smart contract execution is identical to the Mainnet, but does not involve real funds. To simulate a production deployment of a malicious contract by an adversary, we deliberately configure the same contracts with different constructor arguments (e.g., replace token symbol's letter with its homograph twin), or submit additional transactions (e.g., deploy a smart contract at a hard-coded EOA address). It effectively simulates the activation of previously dormant malicious functionality in a production deployment.

Here, we provide a high-level overview of five attack patterns integration.

\noindent \textbf{Integration of \attackIV pattern in Tether USD Stablecoin.}\label{sec:usdt-injection}
\emph{Stablecoin} is a fungible token pegged to the market price of a fiat currency (e.g., U.S. dollar). Adopted mainly by crypto exchanges, mainstream stablecoins have very high market capitalizations and daily transaction volumes. Tether USD (USDT), the most popular stablecoin, is an ERC-20 smart contract deployed on Ethereum\footnote{\texttt{0xdAC17F958D2ee523a2206206994597C13D831ec7}}. We integrate the pattern of attack \attackIV into the source code of USDT by adding a seemingly harmless check of the token symbol before each transfer. We test the code by confirming that the transfer routine's functionality remains unchanged. After that, we simulate a production deployment of the code with an invisible modification of the token symbol, which is passed through the constructor. As a result, the smart contract traps user tokens due to the tampered token symbol.

\noindent \textbf{Integration of \attackV pattern in Binance Token.}\label{sec:bnb-injection}
The Binance Token (BNB)\footnote{\texttt{0xB8c77482e45F1F44dE1745F52C74426C631bDD52}} is a popular ERC-20 altcoin with a high market capitalization and daily transaction volume, collateralized by the financial assets of Binance, a large crypto exchange. We integrate the pattern of attack \attackV into the source code of the BNB token by adding an innocently-looking logging routine, which saves the transfer record in another smart contract. In the test, the code performs logging as expected. However, in the final deployment, the owner replaces one letter in the logging function ICC header with a homograph twin. The log call throws an exception ensuing the failure of fund transfer to users.

\noindent \textbf{Integration of \attackI pattern in ChainLink Token.}\label{sec:link-injection}
A \emph{blockchain oracle} is a service that delivers a reliable outside information into the context of a smart contract. Collateralized by its business assets, ChainLink issues an ERC-20 token with the symbol LINK\footnote{\texttt{0x514910771af9ca656af840dff83e8264ecf986ca}}, in the source code of which we integrate the pattern of attack \attackI. In this token, we use a special user role, the \textit{VIP user}, who can transfer funds at any time, whilst the remaining users can only transfer funds after a pre-determined deadline. The test run does not reveal any issues, but in the production deployment, the malicious smart contract owner mines a similar public address with the same EIP-55 checksum as in the legitimate VIP user address, and saves this address in the smart contract. As a result, the VIP user, who does not recognize the address falsification, will fail to transfer funds from the smart contract.

\noindent \textbf{Integration of \attackVI pattern in Bitfinex Token.}\label{sec:leo-injection}
The Bitfinex LEO token, also known as the UNUS SED LEO\footnote{\texttt{0x2af5d2ad76741191d15dfe7bf6ac92d4bd912ca3}}, is backed by the assets of the Bitfinex crypto exchange. In this token, an auxiliary helper smart contract is used by the attacker for purported protection against transfer flood (i.e., performing too many small transfers by one user). This smart contract uses a homograph substitution of the ICC header of the expected flood-checking function. However, because of the homograph substitution, a wrong function in the auxiliary smart contract is called, which causes an unexpected failure of fund transfer.

\noindent \textbf{Hybrid Social Engineering Attack Pattern Integration in CryptoKitties.}\label{sec:kitties-injection}
The ERC-721 standard is used for non-fungible (i.e., unique) Ethereum tokens, such as collectibles, games, deeds, etc. The CryptoKitties collectible game is one of the most popular ERC-721 tokens\footnote{\texttt{0x06012c8cf97bead5deae237070f9587f8e7a266d}}. For this contract, we use a combination of techniques from attacks \attackI and \attackII. Specifically, the \attackI component involves a manual change of the fee collector by the attacker. The \attackII component deploys a non-payable smart contract at an EOA address, resulting in transaction reversal. Akin to the four previous attacks on ERC-20 tokens, this social engineering exploitation also does not reveal itself during testing: only in the production environment, when the owner deploys the non-payable contract, the malicious logic enables.

%% file: evaluation.tex
\section{Evaluation and Analysis}\label{sec:evaluation}

In this section, we attempt to project the social engineering attacks onto all deployed open source smart contracts and estimate the overall danger of the attacks.

\subsection{Methodology}\label{section:methodology}

As demonstrated in Sections~\ref{sec:threat-model} and \ref{sec:se-attacks}, the detection of social engineering attacks is impossible in a fully-automated manner because human assessment is necessary for understanding semantics of smart contracts. However, manual detection of social engineering attacks requires a laborious effort, such as inspecting the source code with a hex viewer, generating ICC selectors, etc. To address this dichotomy, we develop an automated tool that selects a potential subset of candidates from a given set of smart contracts for further manual analysis. Using this hybrid approach, we manage to filter out over 95.4\% of all the candidates. Then, we manually inspect each of the suspected smart contracts and classify them into three categories: non-exploitable, syntactically matching, and semantically exploitable. Finally, we share our findings with security experts from seven leading smart contract security firms and ask them to share their opinions about the attacks in the form of an online survey.

\subsection{Automated Detection}

\begin{figure}[t]
    \centering
    \includegraphics[width=\linewidth]{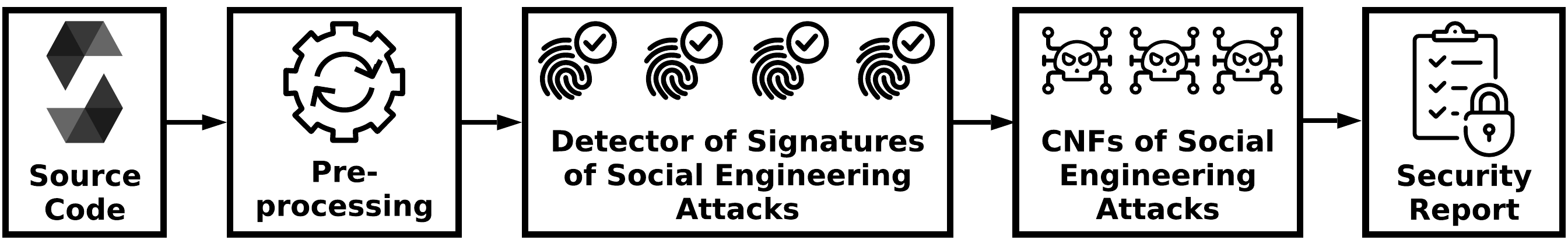}
    \caption{Automated detection of potential social engineering attacks, in which atomic signatures are combined to match an attack profile for each attack in the form of CNF.}
    \label{fig:vcddetector-design}
\end{figure}

A specific feature of all social engineering attacks is that their deception mechanisms are located only in the source code, and therefore undetectable in the bytecode. As a result, we consider the source code of a smart contract as an input. Fig.~\ref{fig:vcddetector-design} illustrates the operation of our automated filter, which uses a double-layer detection, i.e., search for atomic signatures (attack markers) followed by logic processing of these signatures to match specific attacks. First, we preprocess the source codes by parsing multi-file contracts embedded in JSON objects, removing all non-Solidity smart contracts, erasing all the comments, and discarding smart contracts that are duplicates of the previously processed ones. Then, we feed the source codes into a set of signature detectors. Each signature detector utilizes text search and regular expression matching to identify specific \emph{markers} in the source codes. For example, a fund transfer routine can be represented in the source code by either of the three markers: a) the \ttt{transfer} routine; b) the \ttt{send} routine; or c) the \emph{call with value} procedure. These markers are then combined into a signature for detecting a fund transfer. Based on the signatures, we generate social engineering attack detection rules in a conjunctive normal form (CNF) by concatenating a sequence of signatures. The list of all attacks' signatures and CNFs is available in Appendix A. We implement the smart contract scanner using Python, ethereum.utils, and Web3.py, and we publish the source code of the tool at \emph{\url{https://github.com/nick-ivanov/esead}}.

It is worth noting that we do not attempt to detect the proposed social engineering attacks using traditional smart contract vulnerability scanners (e.g., Securify, Sereum, etc.), because these tools by design assume a threat model in which a smart contract is the attack target. The only publicly available tool that fits the threat model of the proposed attacks is HoneyBadger\footnote{https://github.com/christoftorres/HoneyBadger}. However, HoneyBadger is designed to detect Ethereum honeypots --- the type of attack excluded from our study due to its limited audience of targeted victims. Therefore, none of the existing tools is capable to identify the proposed social engineering attacks.

\subsection{Potentially Exploitable Smart Contracts}

Attacks exploiting smart contract code vulnerabilities (e.g., reentrancy or integer overflow) can be detected via automated analysis of bytecode, source code, or transaction history of a smart contract. However, this information is insufficient to identify social engineering attacks with satisfying certainty. For example, consider transaction \ttt{0xc215b9356db58ce05412439f49a842f8a3abe6c179} \ttt{2ff8f2c3ee425c3501023c}, through which the sender paid around \$5 million in gas fees: the context of this transaction cannot be known without a testimony from the sender. Our exhaustive effort to find any existing reports of social engineering attacks in the wild have not yielded any results beyond the cases of honeypot exploitations. Therefore, until the emergence of reports from victims, we can only discuss the potential of the social engineering attacks in the  real-world smart contracts. 

To shed light on the potential existence of social engineering attacks in Ethereum, we collect all available open-source smart contracts from Etherscan\footnote{https://etherscan.io/}, 85,656 unique smart contracts in total, including 73,933 in Mainnet, 8,297 in Ropsten testnet, and 3,426 in Kovan testnet. Table~\ref{tab:smalltable} shows the breakdown of the 3,855 detected candidates, which can potentially deliver social engineering attacks. Then, we perform a manual analysis of all the 3,855 suspicious cases to remove 2,375 non-exploitable smart contracts, and subdivide the remaining 1,480 contracts into 453 \emph{syntactically matching} (but not exploitable) and 1,027 \emph{semantically exploitable} contracts. An example of a non-exploitable contract\footnote{For example, \texttt{0xa62bf7c97c4270882a9278c6f9d684d30e242e03}.} would be the one with a suspicious transfer isolated from critical instructions by a mutually-exclusive \ttt{if-else} branching. Next, we elaborate on how we identify syntactically matching and semantically exploitable contracts, as well as their implications.

\subsubsection{Syntactically Matching Contracts}
A syntactically matching smart contract fits the profile of one of the social engineering attacks (\attackI ... \attackVI), but does not exhibit a deception capability necessary for fooling the victim. For example, smart contract \ttt{0xe5b288da8fb70cd} \ttt{58ab240f71610576657308762} fits the \attackII case because it has a hard-coded fee-collecting EOA address. However, the manual examination of the smart contract reveals that this address is \ttt{0xfeefeefeefee} \ttt{feefeefeefeefeefeefeefeefeef}. Obviously, it is extremely unlikely that someone owns an account that can deploy a smart contract at this address.

Another example of a syntactically matching smart contract is the smart contract called MyMillions\footnote{Deployed at \texttt{0xbBbeCd6ee8D2972B4905634177C56ad73F226276}.}, in which a fee transfer is sharing the call stack of the same transaction with another transfer, while the fee address is both pre-initialized and can be changed, which matches both \attackI and \attackII attacks. However, the manual analysis of this contract reveals that the double transfer occurs in the the function \ttt{buyFactory}, which is an engagement function (i.e., the function that the client calls to participate in the scheme of the smart contract). If this function fails due to the attack, the client deposit will never happen, and therefore this attack will not bring any gain for the attacker. Since semantics of smart contracts vary, only a human can definitely identify engagement and resolution functions.

\subsubsection{Semantically Exploitable Contracts}
A semantically exploitable smart contract not only matches the profile of one of the social engineering attacks, but it also has the deception capability. It indicates that this type of contracts is actually exploitable. A deception capability is an introspective measure characterized by a substantial chance for a contract user to misconstrue the logic of the smart contract, leading to a potential execution of one of the social engineering attacks. The introspective nature of deception capability requires a human to reason about deceptiveness, leading us to manually analyze the source codes of all the 3,855 automatically selected suspected source codes, taking around 140 person-hours in total.

As an example of semantic exploitability, our analysis reveals 34 smart contracts where a comparison with an empty string literal precedes a critical operation, such as the one shown in Fig.~\ref{fig:emptystring}. One way such a contract can be used as a carrier of attack \attackIV is through the use of a zero-width space (Unicode \ttt{U+200B}), which appears as an empty string in many popular text editors (e.g., VS Code). Although none of the suspected 34 contracts have an actual zero-width space, a redeployment of the same contract can be used to launch the social engineering attack \attackIV.

Another interesting exploitable example of attack \attackIV can be found at \ttt{0xf5615138A7f2605e382375fa33Ab368661e017ff}. This smart contract implements a personal smart contract scheme, which implies that each user of the scheme has an individual deployment of the same smart contract, sometimes referred to as a ``wallet''. The contract uses a homograph symbol in a hashmap key, which leads to the inability to withdraw previously deposited funds. Although the contract has an obvious deception capability, neither code nor transaction log could definitely determine the contract's maliciousness. In other words, the homograph substitution of the map key may indicate a malice or a mere typo.

Another peculiar example of a semantically exploitable Address Manipulation attack is the game called \emph{JigsawGames2}\footnote{Deployed at \texttt{0x2C7Bc39B1B0C9Fdf200fd30C74C0a9a41C2C7047}.}. In this contract, the resolution function \ttt{sellEggs} contains a fee transfer alongside with the user reward transfer, which allows the attacker to block the user from getting the prize by making the fee address non-payable via attack \attackI or \attackII techniques. The contract does not implement any self-destruction or deprecation functionality, posing a challenge for the attacker who needs to acquire the funds trapped in the contract. Coincidentally, this smart contract also charges a developer fee in the engagement function \ttt{buyEggs}. In this case, the attacker can create a fake player, and make the fee address payable by  calling \ttt{buyEggs} function multiple times using the fake player until the contract balance is drained through multiple fee transfers. This example shows that smart contract owners often have multiple indirect ways of stealing funds from smart contracts.

\begin{figure}[t]
    \centering
\begin{lstlisting}[language=Solidity]
if (!compareStr(userGlobal.referrer, "")) {
  ...
  userRoundMapping[rid][referrerAddr].inviteAmount++;
}
\end{lstlisting}
    \caption{Empty string comparison in contract \ttt{0x61394198ee6cbe2d6ad603d52c10fba3237202ef}.}
    \label{fig:emptystring}
\end{figure}

\subsection{Observations}

While performing a manual analysis of 3,855 suspected smart contracts, we gathered some interesting observations, which are relevant within a broader discussion about social engineering attacks in Ethereum.

\noindent \textbf{Observation 1 [Multiple versions of the same code]:}
It is well-known that a vast majority of smart contracts reuse secure patterns, modifiers, and abstract classes from the OpenZeppelin Contracts library. However, despite the fact that we remove all duplicate smart contracts during the pre-processing stage, our manual analysis of the suspected smart contracts reveals a significant number of large contract clusters, in which a custom code is reused with slight modifications. Such clusters of reused custom code patterns are also widely presented in the semantically exploitable set, which demonstrates that code reuse is prevalent in smart contracts,  leading to the dissemination  of insecure patterns.

\noindent \textbf{Observation 2 [No evidence of testnet experimentation with social engineering attacks]:} In pursuit of early signs of experimentation with social engineering attack patterns, we supplement our dataset with open-source contracts from two testnets --- Ropsten and Kovan. Our initial hypothesis was that the first experimental exploitations of social engineering attacks may prevail at testnets first. However, compared to Mainnet, in which 937 out of 3,165 suspected contracts are semantically exploitable (29.6\%), in Ropsten this is 11.9\%, and in Kovan it is 16.0\%. Thus, the testnets exhibit reduced probability of encountering semantically exploitable social engineering contracts.

\begin{table}[t]
    \centering
    \caption{Analysis results of 85,656 smart contracts.}
    \label{tab:smalltable}
    
    \begin{tabular}{|c|c|c|c|}
    \hline
    \multirow{2}{*}{\textbf{\small Attack}} &
    \textbf{\small Non-} & \textbf{\small Syntactically} & \textbf{\small Semantically} \\
        & \textbf{\small exploitable} & \textbf{\small matching} & \textbf{\small exploitable} \\
    \hline
    \attackI
    & 561 & 230 & 636 \\
    \hline
    
    \attackII
    & 213 & 100 & 341 \\
    \hline
    
    \attackIII 
    & 1,515 & 0 & 0 \\
    \hline
    
    \attackIV
    & 86 & 123 & 50 \\
    \hline
    
    \attackV
    & 0 & 0 & 0 \\
    \hline
    
    \attackVI
    & 0 & 0 & 0 \\
    \hline
    
    \textbf{Total:} & 
    2,375 &  453 & 1,027 \\
    \hline
    \end{tabular}
\end{table}

\begin{figure}[t]
\centering
\begin{subfigure}{\linewidth}
  \centering
    \includegraphics[width=2.5in]{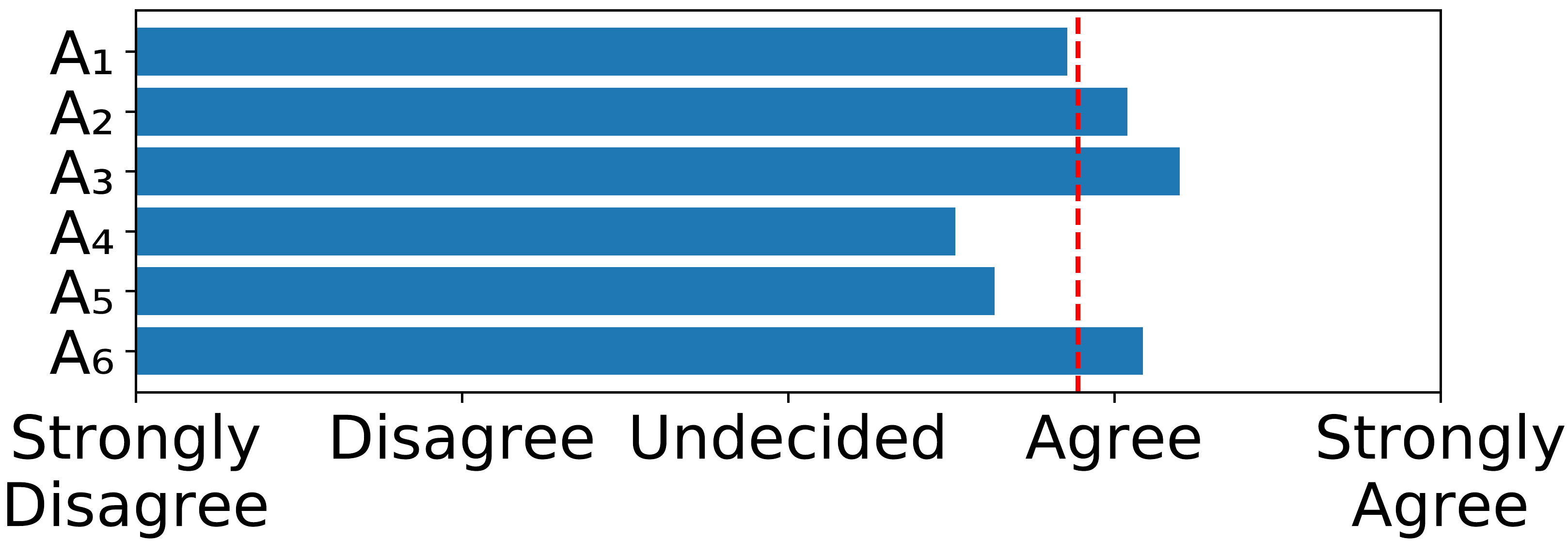}
  \caption{Could this attack be dangerous to your customers?}
  \label{fig:sur1}
\end{subfigure}
\hspace{5pt}
\begin{subfigure}{\linewidth}
  \centering
    \includegraphics[width=2.5in]{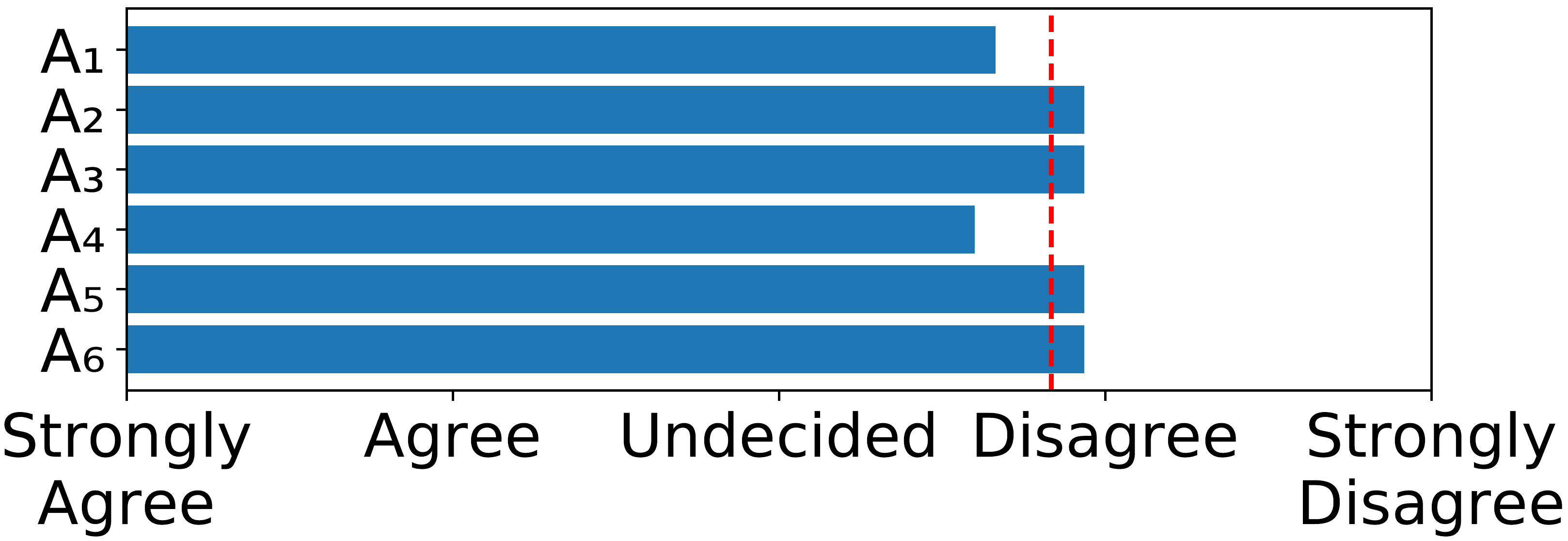}
  \caption{Do you think the attack can be discovered by human users?}
  \label{fig:sur3}
\end{subfigure}
\caption{Average survey results from seven smart contract auditing firms. The red vertical line represents the average value of the six attacks.}
\label{fig:sur}
\end{figure}

\subsection{Survey of Auditing Firms}
\label{sec:firms}
To further evaluate the proposed attacks, we send surveys consisting of two questions shown in Fig.~\ref{fig:sur} to the following seven smart contract firms (listed alphabetically): \mbox{Audithor}, \mbox{CertiK}, \mbox{CoinFabrik}, \mbox{ConsenSys}, \mbox{Dedaub}, \mbox{Trail of Bits}, and one company that elected to be anonymous. The responses were provided by actual smart contract developers and security auditors from each of the firms (one participant from each company), including 6 males and 1 female professionals. Fig.~\ref{fig:sur} represents the answers from the experts regarding the six social engineering attacks. The vertical red lines represent the averages of responses with respect to all the six attacks. The results of the survey demonstrate that the experts agree that the social engineering attacks can cause damage to their customers. Also, the experts believe that the social engineering attacks are unlikely to be discovered by a human user.

%% file: recommendations.tex
\section{Security Recommendations}\label{sec:security-rec}

In Section~\ref{sec:evaluation}, we demonstrate that even if all the syntactic patterns in a smart contract correctly match one of the social engineering attacks, only 1,027 contracts out of total 3,855 are actually exploitable, which is less than 27\%. Corroborating our finding, Zhou et al.~\cite{zhou2020ever} demonstrate that the attempt to detect Ethereum honeypots by Torres et al.~\cite{torres2019art} in a fully-automated manner produces a large number of false negative and false positive results. Therefore, the defense against social engineering attacks should involve human auditing. To account for this characteristic of social engineering attacks, we develop a list of recommendations for people considering engagement with a smart contract, including security auditors verifying safety of smart contracts on behalf of their clients.  These recommendations aim for effective identification and prevention of social engineering attacks with minimal effort. 

\noindent \textbf{Recommendation 1 [Beware of address change]:} \emph{To prevent \attackI, smart contract users should not engage in a contract which allows to change the address that is a transfer recipient within the call stack of a critical operation.}
Our analysis finds many smart contracts with such patterns in the wild, but none of them exhibit a malicious intent or have a suspicious history. However, it grants a potential backdoor for the owner to block critical operations, e.g., fund withdrawals.

\noindent \textbf{Recommendation 2 [Check EOAs for outgoing transactions]:} \emph{To prevent \attackII, smart contract users should verify that all hard-coded EOAs have at least one outgoing transaction}. If the EOA has outgoing transactions (marked as ``OUT'' by Etherscan), it indicates that the smart contract owner knows the private key of the EOA, and it entails that the owner does not know the private key of the account that could deploy a smart contract at this address. In fact, the probability that someone knows the private key of an EOA \emph{and} the private key of the account for deploying a contract at the same address equals to the probability of a 160-bit hash collision because each public address is a Keccak256 hash of a public key trimmed to 160 bits.

\noindent \textbf{Recommendation 3 [Avoid visual cognitive bias]:} \emph{To prevent \attackI, smart contract users should never compare addresses visually; text editor search function should be used instead.} In this paper we show that EIP-55 collision bruteforce attacks are easy to carry out. As a result, even slightly modified addresses with unknown associated private keys can be dangerous. Therefore, users should treat all public addresses with suspicion.

\noindent \textbf{Recommendation 4 [Avoid confirmation bias]:} \emph{To prevent \attackIII, smart contract users should never use accounts with all-lowercase EIP-55 checksums for smart contract testing.} Most Ethereum clients, such as Metamask, enforce EIP-55 checksums, so public addresses are always shown in a mixed-capitalization form. Another way to verify an address is to paste it in the search field of Etherscan, which also enforces EIP-55. If the address is all-lowercase, it might be a part of a social engineering scheme, and thus the contract should undergo additional scrutiny.

\noindent \textbf{Recommendation 5 [Do not trust string comparison]:} \emph{To prevent \attackIV, smart contract users should not engage in a smart contract that uses string comparison to determine a transfer or another critical operation.}
If a text comparison involves two immutable values, e.g., constant and string literal, it is essentially a tautology, and is indicative of a derelict smart contract. However, one way to carry out attack \attackIV is to mimic a tautology, as is shown in Fig.~\ref{fig:badchoice20}. Either way, a critical operation determined by a string comparison should be treated with caution.

\noindent \textbf{Recommendation 6 [Verify ICC selectors]:} \emph{To prevent \attackV and \attackVI, smart contract users should verify the arguments of \ttt{call()} and \ttt{delegatecall()} with a hex viewer.} Smart contract users and auditors cannot see selectors associated with functions and arguments of \ttt{call()}/\ttt{delegatecall()} while examining the Solidity code, since these selectors are computed at the compile time. If the parameters of \ttt{call()} or \ttt{delegatecall()} include a string literal, we recommend to compile both the calling and the callable contracts with \ttt{--asm} or \ttt{--ir} options to verify that the selectors of functions match. If the parameters are mutable variables, the contract cannot be treated as safe.

%% file: relatedwork.tex
\section{Related Work}

The study of social engineering attacks in Ethereum is limited to honeypots --- deceptive smart contracts targeting users who attempt to exploit known vulnerabilities of smart contracts. Torres et al.~\cite{torres2019art} present a taxonomy of \emph{honeypots}, while Zhou et al.~\cite{zhou2020ever} later discover 51 previously undetected honeypots. Although Ethereum honeypots is definitely a subclass of social engineering attacks, these contracts are harmless for ordinary users, as their potential victims are opportunistic malicious players.

The type of social engineering attacks we discovered in this paper have been known outside of the blockchain domain. Fu et al.~\cite{fu2006methodology} present a methodology for defending against such attacks, and develop a Unicode character similarity list and attack detection tool, IDN-SecuChecker. Holgers et al.~\cite{holgers2006cutting} conduct a measurement study of IDN homograph attacks, which shows their real-world impact. However, our research is the first to successfully apply these techniques to Ethereum smart contracts.

Email/URL phishing and Ethereum social engineering attacks both target human cognitive biases. Phishing attacks have been thoroughly studied in recent years~\cite{ho2019detecting,van2019cognitive,hu2018end,szurdi2014long,cranor2007phinding,neupane2014neural,neupane2015multi}. However, the unique characteristics of smart contracts, such as open execution, fee-charging transactions, and non-interactive properties, make the design of their social engineering attacks significantly different from traditional phishing attacks.

%% file: conclusion.tex
\section{Conclusion}

This work zeroes in on a largely overlooked class of social engineering attacks in Ethereum smart contracts. These attacks exploit human cognitive biases as new attacking vectors. We identified these biases and developed six zero-day social engineering attacks. By embedding most of these attacks into existing popular tokens, we demonstrated that the attacks have the potential to victimize a large group of normal users. Moreover, the attacks remain dormant during testing and only activate after a production deployment. We worked with seven smart contract security firms and confirmed that the attacks are indeed dangerous and evasive. Our analysis reveals 1,027 existing smart contracts that can potentially carry out social engineering attacks. By open-sourcing our analysis tools and benchmark datasets, we invite further research exploration of this emerging topic.

%% file: acknowledgement.tex
\section*{Acknowledgement}
We would like to thank Jon Gorman and other anonymous reviewers for providing valuable feedback on our
work. This work was supported in part by National Science Foundation grants CNS1950171 and CNS1949753.

%% file: appendix.tex
\section*{Appendix}
\section*{A. ~~~Attack Signatures}\label{appendix:vcddetector-signatures}
Table~\ref{tab:vcddetector-signatures} provides a full list of signatures that we use to detect potential social engineering attacks, based on which we generate the CNF detection rule for each of the six social engineering attacks, which are defined as follows:
$$
\begin{array}{l}
CNF(\attackI) = S_{1} \wedge (S_{2} \vee S_{3} \vee S_{4}) \wedge S_{5} \\
CNF(\attackII) = (S_{2} \vee S_{3} \vee S_{4}) \wedge S_{5} \wedge S_{6} \wedge (S_{7} \vee S_{8}) \wedge S_{9} \\ 
CNF(\attackIII) = S_{5}  \wedge S_{10} \wedge (S_{11} \vee S_{12} \vee S_{13} \vee S_{14}) \wedge S_{15} \\
CNF(\attackIV) = S_{5} \wedge (S_{11} \vee S_{12} \vee S_{13} \vee S_{14}) \wedge S_{16} \wedge (S_{17} \vee S_{18}) \\
CNF(\attackV) = S_{5} \wedge (S_{11} \vee S_{12} \vee S_{13} \vee S_{14}) \wedge (S_{19} \vee S_{20}) \wedge S_{21}  \\
CNF(\attackVI) = S_{5} \wedge (S_{11} \vee S_{12} \vee S_{13} \vee S_{14}) \wedge (S_{19} \vee S_{20}) \wedge S_{21} \wedge S_{22}
\end{array}
$$
\begin{table*}[t]
    \caption{The full list of signatures used for automated detection of the six social engineering attacks.}
    \label{tab:vcddetector-signatures}
    
    \centering
        \begin{tabular}{ccl}
          \toprule[1.5pt]
          \head{Symbol} & \head{Social Engineering Signature} & \head{Matching Attacks} \\
          \midrule
          $S_{1}$ & Non-constructor public or external function that alters an address variable  & \attackI  \\
          $S_{2}$ & Ether transfer with another Ether transfer in the call stack of the same transaction & \attackI, \attackII  \\
          $S_{3}$ & Ether transfer with call-with-value statement in the call stack of the same transaction & \attackI, \attackII  \\
          $S_{4}$ & Ether transfer with a token transfer in the call stack of the same transaction & \attackI, \attackII  \\
          $S_{5}$ & Smart contract has a payable function & \attackI, \attackII, \attackIII, \attackIV, \attackV, \attackVI \\
          $S_{6}$ & \ttt{emit} instruction inside a call stack of a payable function & \attackII \\
          $S_{7}$ & Constant variable with \ttt{address} type and a hard-coded value & \attackII \\
          $S_{8}$ & Non-constant variable with \ttt{address} type and a hard-coded value & \attackII \\
          $S_{9}$ & Ether transfer to an address variable initialized with a hard-coded value & \attackII \\
          $S_{10}$ & Hard-coded \ttt{bytes32} value & \attackIII \\
          $S_{11}$ & Ether transfer inside a branching arm & \attackIII, \attackIV, \attackV, \attackVI \\
          $S_{12}$ & Token transfer inside a branching arm & \attackIII, \attackIV, \attackV, \attackVI \\
          $S_{13}$ & Ether transfer with a \ttt{require} statement in the call stack of the same transaction & \attackIII, \attackIV, \attackV, \attackVI \\
          $S_{14}$ & Token transfer with a \ttt{require} statement in the call stack of the same transaction & \attackIII, \attackIV, \attackV, \attackVI \\
          $S_{15}$ & \ttt{bytes32} value inside a branching condition &  \attackIII \\
          $S_{16}$ & Comparison of Keccak256 hash values & \attackIV \\
          $S_{17}$ & String literal as part of a branching condition & \attackIV \\
          $S_{18}$ & String literal as part of a \ttt{require} statement & \attackIV \\
          $S_{19}$ & Ether transfer with \ttt{call} or \ttt{delegatecall} statement in the call stack of the same transaction & \attackV, \attackVI \\
          $S_{20}$ & Token transfer with \ttt{call} or \ttt{delegatecall} statement in the call stack of the same transaction & \attackV, \attackVI \\
          $S_{21}$ & String literal with a non-ASCII symbol somewhere in the contract & \attackV, \attackVI \\
          $S_{22}$ & ICC status is used in a \ttt{require} statement & \attackVI \\
        \bottomrule[1.5pt]
        \end{tabular}
\end{table*}

\section*{B. ~~~Address Miner}

We develop an address miner to mine Ethereum addresses with all lower-case EIP-55 checksums. Table~\ref{tab:lowercase-eip55} shows five sample addresses. Such addresses can be used in the \attackIII attack. 

\begin{table*}[t]
\small
\caption{Sample lowercase EIP-55-compliant addresses.}
\begin{tabular}{|c|c|c|}
\hline
\textbf{EIP-55-compliant Lowercase Address} & \textbf{Private Key of the Account} &                                                                                                \textbf{Mining Time (ms)} \\ \hline

\texttt{0x47aa51fd5a98e155623202944c44f414a7205a46}  & \begin{tabular}[c]{@{}c@{}}\texttt{bed6ad86fa57efe205abdcda885b3010}\\ \texttt{7b1a75d6196b271d4785cd3ed66c8d5d}\end{tabular} & 6,822 \\ 
 \hline

\texttt{0x8310561552fa9569337d53493c6a5a8991894072}  & \begin{tabular}[c]{@{}c@{}}\texttt{4856d3e9c032724eca42a5fd48e99dc5}\\ \texttt{b77cb5be96ca68eb9e03511257999e61}\end{tabular} & 3,137 \\ 
 \hline

\texttt{0x2797a2c394686d33da258c7de6206617c398605e}  & \begin{tabular}[c]{@{}c@{}}\texttt{1321d554cddf1b756e8d15cba0a33fb4}\\ \texttt{e84b95119acf8e267f7505f29f652020}\end{tabular} & 460 \\ 
 \hline
 
\texttt{0x596443674c431e7da447803ef94a7e52cfd71169}  & \begin{tabular}[c]{@{}c@{}}\texttt{1265ca0334308e3dfb2ddd9a7eb466aa}\\ \texttt{488a863671e6ad6290d93383489159d1}\end{tabular} & 1,954 \\ 
 \hline
 
\texttt{0x52206f3a3b80212898760a6ae124474183b30612}  & \begin{tabular}[c]{@{}c@{}}\texttt{a532795660fbb9ccb5f3862e102f1968}\\ \texttt{0a5def583aea24a2875de7f1dd6c8298}\end{tabular} & 266 \\ 
 \hline
 
\texttt{0xc71c3eec3aa44e7746725fc771b8b821419e4360}  & \begin{tabular}[c]{@{}c@{}}\texttt{3b1b3a32d73bd32f837440cd0469a801}\\ \texttt{0fa6f3e02358ffeb76c95454ee2a0e36}\end{tabular} & 4,896 \\ 
 \hline

\end{tabular}

\label{tab:lowercase-eip55}
\end{table*}

\section*{C. ~~~Integrating Social Engineering Attack Patterns into Existing Tokens}\label{appendix:exploitation}

\noindent\textbf{\attackIV Attack Pattern Integration in USDT.} In Fig.~\ref{fig:usdt-vcd-injection}, we show that without changing the logic of the smart contract, the \attackIV social engineering attack pattern can be integrated into the Tether stablecoin source code.
Specifically, in the Tether USD token, we add a seemingly harmless check of the token symbol within the ERC-20 \ttt{transfer}. The evasive test deployment uses all-Latin characters for token symbols, whereas the malicious smart contract is deployed by passing to the constructor a token symbol with unnoticeable substitution of one character, which leads to the failure of the fund transfer.

\begin{figure}[t]
    \centering
\begin{lstlisting}[language=Solidity]
if(keccak256(abi.encode(symbol)) == keccak256(abi.encode("USDT"))) {
  return super.transfer(_to, _value);
}
\end{lstlisting}
    \caption{Integration of the \attackIV attack pattern into the \ttt{transfer} ERC-20 method of Tether stablecoin source code.}
    \label{fig:usdt-vcd-injection}
\end{figure}

\noindent\textbf{\attackV Attack Pattern Integration in BNB.} Fig.~\ref{fig:bnb-vcd-injection} shows an integration of the \attackV attack pattern into the Binance exchange token source code. Fig.~\ref{fig:bnb-vcd-helper} shows the helper class for the \attackV attack in the Binance Token. In the \ttt{transfer} method (Fig.~\ref{fig:bnb-vcd-injection}), we insert a logging routine, which saves the transfer record in a consolidated database in another smart contract (Fig.~\ref{fig:bnb-vcd-helper}). In a test deployment, the code performs logging as expected. However, in the final deployment, the owner replaces one letter in the logging function header with a homograph twin, e.g., the second letter ``o'' with the identically-looking Cyrillic letter. The log call (Fig.~\ref{fig:bnb-vcd-injection}, line 3) throws an exception and the transfer fails.

\begin{figure}[t]
    \centering
\begin{lstlisting}[language=Solidity]
address consolidatedDBAddress = 
  0x51Db8896d6bD64385C5785Df0685cc4C24F01F0f;
bytes memory payload = abi.encodeWithSignature("logVolume(address,uint256)", _to, _value);
bool success = address(consolidatedDBAddress).call(payload);
if(success) {
  balanceOf[msg.sender] = SafeMath.safeSub(balanceOf[msg.sender], _value);
  balanceOf[_to] = SafeMath.safeAdd(balanceOf[_to], _value);
  Transfer(msg.sender, _to, _value);
}
\end{lstlisting}
    \caption{Integration of the \attackV attack pattern into the \ttt{transfer} method of the Binance exchange token source code.}
    \label{fig:bnb-vcd-injection}
\end{figure}

\begin{figure}[t]
    \centering
\begin{lstlisting}[language=Solidity]
function logVolume(address client, uint256 amount) public {
  require(msg.sender==authorizedCallerSmartContract);
  clientVolumes[client] += amount;
}
\end{lstlisting}
    \caption{Function \ttt{logVolume} in the helper contract used for the \attackV attack in the Binance exchange token.}
    \label{fig:bnb-vcd-helper}
\end{figure}

\noindent\textbf{\attackI Attack Pattern Integration in LINK.} In this token, the malicious smart contract owner mines a similar public address with the same EIP-55 checksum as in the original address, and initializes \ttt{vipClient} via the constructor (Fig.~\ref{fig:link-vcd-injection}, line 5). As a result, the VIP user, who does not recognize the address falsification, will fail to transfer funds.

\begin{figure}[t]
    \centering
\begin{lstlisting}[language=Solidity]
function LinkToken(address vc) public
{
  balances[msg.sender] = totalSupply;
  transferAllowedAfterBlock = block.number + (2 * 365 * 24 * 60 * 6);
  vipClient = vc;
 owner = msg.sender;
}
...
function transfer(address _to, uint _value) public
validRecipient(_to) returns (bool success) {
  if(block.number > transferAllowedAfterBlock || msg.sender == vipClient || msg.sender == owner) {
    return super.transfer(_to, _value);
  }
}
\end{lstlisting}
    \caption{Integration of the \attackI attack pattern into the \ttt{transfer} method of the ChainLink oracle token source code.}
    \label{fig:link-vcd-injection}
\end{figure}

\noindent\textbf{\attackVI Attack Pattern Integration in LEO.} Fig.~\ref{fig:leo-vcd-injection} shows an integration of the \textit{\attackVI} attack pattern into the token's source code. Fig.~\ref{fig:leo-vcd-helper} shows the helper class for the \attackVI attack in the Bitfinex Token. In this token, a helper smart contract is used by the attacker for purported protection against transfer flood, i.e., performing too many small transfers by one user. The smart contract (see Fig.~\ref{fig:leo-vcd-helper}) has two functions, \ttt{logAndCheck}, and seemingly unrelated and benign \ttt{onCurve34906537}. However, the latter function is the one called by the token smart contract due to homograph substitution of several symbols in the \ttt{call} argument. Unlike in the \attackV attack against the BNB token, the attack \attackVI does not require to change the original ICC header before the production deployment. Instead, the contract owner simply changes the value of \ttt{extraFeaturesEnabled} flag to activate the attack.

\begin{figure}[t]
    \centering
\begin{lstlisting}[language=Solidity]
address floodProtectionSC = 
  0x5B38C7add838EfFF53412C71E9efF5c182c6b407;
bytes memory payload = abi.encodeWithSignature("logAndCheck(address)", msg.sender);
(bool succ, bytes memory result) = address(floodProtectionSC).call(payload);
require(succ);
if(abi.decode(result, (bool)) == true) {
  doTransfer(msg.sender, _to, _amount);
  return true;
}
\end{lstlisting}
    \caption{Integration of the \attackVI attack pattern into the \ttt{transfer} ERC-20 call of the Bitfinex LEO source code.}
    \label{fig:leo-vcd-injection}
\end{figure}

\begin{figure}[t]
    \centering
\begin{lstlisting}[language=Solidity]
function onCurve34906537(address) public view returns (bool) {
  if(extraFeaturesEnabled) {
    return true;
  } 
  return false;
}
function logAndCheck(address client) public returns (bool) {
  require(msg.sender==authorizedCallerSmartContract);
  calls[client] += 1;
  return true;
}
\end{lstlisting}
    \caption{Function \ttt{onCurve34906537} is called instead of \ttt{logAndCheck} in the Helper contract, which is used for the \textit{\attackVI} attack in the Bitfinex LEO token.}
    \label{fig:leo-vcd-helper}
\end{figure}

\begin{figure}[t]
    \centering
\begin{lstlisting}[language=Solidity]
address public fee_collector = 
  0xce02be9dfc4c68bae86a0bdf1bab68de77bb0d8d;
function withdrawBalance() external onlyCEO {
  uint256 balance = this.balance;
  uint256 subtractFees = (pregnantKitties + 1) * autoBirthFee;
  if (balance > subtractFees) {
    fee_collector.transfer(subtractFees);
    cfoAddress.send(balance - subtractFees);
  }
}
\end{lstlisting}
    \caption{A hybrid \attackI + \attackII attack pattern integrated into the \ttt{withdrawBalance} function of the CryptoKitties ERC-721 collectible source code.}
    \label{fig:kitty-vcd-injection}
\end{figure}

\noindent\textbf{Hybrid Attack Pattern Integration in CK.} Fig.~\ref{fig:kitty-vcd-injection} shows an integration of the hybrid \emph{\attackI/\attackII} attack pattern into the CryptoKitties ERC-721 collectible source code. The CryptoKitties smart contract can accept and withdraw Ether. In the function \ttt{withdrawBalance} (see Fig.~\ref{fig:kitty-vcd-injection}), \ttt{send} is preceded by a seemingly safe and reasonable fee collection. This arrangement works impeccably during the testing. However, after the production deployment, the owner of the contract deploys a non-payable smart contract at the address stored in \ttt{fee\_collector}: such a substitution is possible because the address has been pre-calculated in advance as described in Section~\ref{subsec:attackII}.